\documentclass[aip,jap,reprint]{revtex4-1}

\usepackage{amsmath}
\usepackage[utf8]{inputenc}
\usepackage{academicons}
\usepackage{xcolor}
\usepackage{graphics}
\usepackage{graphicx}
\usepackage[T1]{fontenc} 

\usepackage[breaklinks=true]{hyperref}
\usepackage{tikz}
\usepackage[all]{hypcap}
\usepackage{orcidlink}
\definecolor{JAPblue}{rgb}{0, 0.682, 0.937}
\hypersetup{
	colorlinks = true,
	linkcolor = {JAPblue},
	citecolor = {JAPblue},
	urlcolor = {JAPblue},
}
\usepackage{amsfonts}

\usepackage{upgreek}

\begin{document}

\title{X-ray thermal diffuse scattering as a texture-robust temperature diagnostic for dynamically compressed solids}

\author{P.G. Heighway~\orcidlink{0000-0001-6221-0650}}\email{patrick.heighway@physics.ox.ac.uk}
\affiliation{Department of Physics, Clarendon Laboratory, University of Oxford, Parks Road, Oxford OX1 3PU, UK\looseness=-1}

\author{D.J. Peake\orcidlink{0000-0002-5992-6954}}
\affiliation{Department of Physics, Clarendon Laboratory, University of Oxford, Parks Road, Oxford OX1 3PU, UK\looseness=-1}

\author{T. Stevens~\orcidlink{0009-0006-8355-3509}}
\affiliation{Department of Physics, Clarendon Laboratory, University of Oxford, Parks Road, Oxford OX1 3PU, UK\looseness=-1}

\author{J.S. Wark~\orcidlink{0000-0003-3055-3223}}
\affiliation{Department of Physics, Clarendon Laboratory, University of Oxford, Parks Road, Oxford OX1 3PU, UK\looseness=-1}


\author{B. Albertazzi}\affiliation{Ecole Polytechnique, Palaiseau, Laboratoire pour l'utilisation des lasers intenses (LULI), CNRS UMR 7605 Route de Saclay 91128 PALAISEAU Cedex, France\looseness=-1}

\author{S.J. Ali~\orcidlink{0000-0003-1823-3788}}\affiliation{Lawrence Livermore National Laboratory, Livermore, CA 94550, USA\looseness=-1}

\author{L. Antonelli~\orcidlink{0000-0003-0694-948X}}\affiliation{University of York, School of Physics, Engineering and Technology, Heslington York YO10 5DD, UK\looseness=-1}

\author{M.R. Armstrong~\orcidlink{0000-0003-2375-1491}}\affiliation{Lawrence Livermore National Laboratory, Livermore, CA 94550, USA\looseness=-1}

\author{C. Baehtz~\orcidlink{0000-0003-1480-511X}}\affiliation{Helmholtz-Zentrum Dresden-Rossendorf (HZDR), Bautzner Landstra{\ss}e 400, 01328 Dresden, Germany\looseness=-1}

\author{O.B. Ball~\orcidlink{0000-0002-5215-0153}}\affiliation{SUPA, School of Physics and Astronomy, and Centre for Science at Extreme Conditions, The University of Edinburgh, Edinburgh EH9 3FD, UK\looseness=-1}

% ...

\date{\today}

\begin{abstract}
We present a model of x-ray thermal diffuse scattering (TDS) from a cubic polycrystal with an arbitrary crystallographic texture, based on the classic approach of Warren. We compare the predictions of our model with femtosecond x-ray diffraction patterns obtained from ambient and dynamically compressed rolled copper foils obtained at the High Energy Density (HED) instrument of the European X-Ray Free-Electron Laser (EuXFEL), and find that the texture-aware TDS model yields more accurate results than does the conventional powder model owed to Warren. Nevertheless, we further show that: with sufficient angular detector coverage, the TDS signal is largely unchanged by sample orientation and in all cases strongly resembles the signal from a perfectly random powder; shot-to-shot fluctuations in the TDS signal resulting from grain-sampling statistics are at the percent level, in stark contrast to the fluctuations in the Bragg-peak intensities (which are over an order of magnitude greater); and TDS is largely unchanged even following texture evolution caused by compression-induced plastic deformation. We conclude that TDS is robust against texture variation, making it a flexible temperature diagnostic applicable just as well to off-the-shelf commercial foils as to ideal powders.
\end{abstract}

\maketitle

\section{Introduction}

The search for a robust temperature diagnostic is one of the longest-standing\cite{Kormer1968} and most fundamental problems in the field of dynamic-compression science. Temperature controls myriad physical properties and processes, including polymorphism, plasticity, diffusivity, and conductivity, and is therefore essential to understanding the exotic, high-pressure states of matter generated via shock- or ramp-compression. Modern dynamic-compression experiments are often realized via laser ablation\cite{Hawreliak2011,Suggit2012,Milathianaki2013,Wehrenberg2017,McBride2019,Briggs2019,Sharma2020,Lazicki2021,Pandolfi2022,Singh2024}, in which a high-power pulsed optical laser is focused onto the surface of a solid sample, thereby launching a supersonic shock-wave that leaves in its wake material at several million atmospheres, pressures comparable to those found in terrestrial planetary interiors. To measure the temperature of this short-lived, high-density matter within the nanosecond window before which it depressurizes and subsequently disintegrates presents unique experimental challenges. To overcome them, the dynamic-compression community has been developing several independent thermometric techniques, some of which have only borne fruit within the last couple of decades.

The conventional way to measure the temperature of shock-loaded matter is via pyrometry (i.e., photon emission spectroscopy), whereby the thermal radiation emitted by the rear surface of the sample is fitted to a gray-body spectrum\cite{Hereil2000,Swift2007,Millot2018}. While relatively easy to field, pyrometry has a number of known drawbacks. First, for samples opaque to radiation in the infrared-to-visible range, the temperature being measured is that of the sample's \emph{surface}, rather than its bulk. To maintain the shock pressure at the rear surface long enough for measurable spectra to be recorded, the sample is often tamped by an optically transparent `window'. Even on nanosecond timescales, some heat will be conducted between the sample surface and the window\cite{Grover1974,Anderson1996,Brantley2021}. Moreover, any mismatch in shock impedance between the two layers will launch a secondary shock- or release-wave into the sample, causing further temperature changes. Such surface effects complicate the modeling required to back out meaningful temperatures, but they are not insurmountable\cite{Brantley2021}; the second and more fundamental problem with pyrometry is that, for laser-ablation experiments, the duration of the emission is so short ($\sim10$~ns) and the emitting area so small ($<1$~mm\textsuperscript{2}) that the signal may simply be too dim to yield a sensible temperature. This renders pyrometry unusable for poor emitters at temperatures below a few thousand degrees.

The new generation of alternatives to pyrometry is generally based on the volumetric interaction of the sample with a high-intensity x-ray probe. Arguably the most mature of these techniques -- as applied to dynamically compressed matter, at least -- is measurement of the extended x-ray absorption fine structure (EXAFS)\cite{Yaakobi2004,Yaakobi2005,Ping2013,Torchio2016,Turneaure2022}. Here, one illuminates the sample with a bright, broadband x-ray source whose spectrum straddles a K- or L-edge, and records the absorption spectrum; due to interference between the photoelectron ejected from the central atom and the neighboring atoms arranged periodically around it, the absorption coefficient marginally above the edge exhibits modulations, the depth of which depends on the typical thermal displacement of each atom from its equilibrium position. This technique is exemplified by the recent study of Sio \textit{et al.}~\cite{Sio2023} wherein EXAFS measurements of copper (Cu) ramp-compressed to around 400~GPa at the National Ignition Facility (NIF) were compared directly with synthetic absorption spectra calculated using atomistic configurations from classical and first-principles molecular dynamics (MD) simulations, providing temperature measurements with an error of $\pm20\%$. Ultimately, what one measures with EXAFS is the Debye-Waller factor $M$ (a proxy for the mean-squared atomic displacement), which must be related to the temperature $T$ via a model. When not deduced implicitly using MD simulations as in the aforementioned study, $M(T)$ is typically calculated using the Debye model, which necessarily involves numerous simplifications of the underlying physics. This model-dependence is a limitation common to all thermometric techniques based on measurement of the Debye-Waller factor.

A second approach to temperature diagnosis garnering considerable attention is measurement of the inelastic x-ray scattering (IXS) spectra of x-ray photons interacting with phonons. When a monochromatic photon beam is launched at a crystalline sample and the distribution of scattered photon energies at an angle far from the Bragg peaks recorded, the resultant spectra depend on the typical values of $M$ at the measured scattering angles (and temperatures). For low $M$ -- corresponding physically to low-$\mathbf{q}$ or forward scattering, where scattering from single phonons dominates -- one generally observes two peaks symmetrically displaced about the original photon energy, corresponding to photons that have either emitted or absorbed a phonon; the relative probabilities of these interactions occurring is controlled by detailed balance, and hence the system temperature. IXS in this mode thus provides what is essentially a model-independent thermometer, along with information about the phonon frequencies (and hence Debye temperature $\Theta_D$). For experimental geometries where higher values of $M$ are accessed (e.g., in backscatter), multiphonon scattering dominates, and the recorded spectrum tends to a Gaussian distribution, the width of which is determined by temperature.

The implementation of IXS at the European X-ray Free-Electron Laser (EuXFEL) facility has been pioneered by the group of McBride \textit{et al.}\cite{McBride2018,Descamps2020a,Wollenweber2021}. Presently, the main barrier to realizing IXS-based temperature measurements in dynamically compressed matter is the difficulty accruing enough data to overcome the exceptionally small scattering cross-section, given that resolutions of a few tens of meV are required; even using a monochromated XFEL -- the most spectrally bright light-source available -- at least 100 repeated shots per compression state would likely be required to obtain a single temperature measurement. Such a workflow is now becoming practicable using highly stable, high-repetition-rate drivers like the DiPOLE 100-X laser\cite{Mason2018,Gorman2024}, but IXS is in any case fundamentally incapable of providing \emph{single-shot} temperature measurements using current light-source technology. There also exists an upper limit on the temperature that can be derived from forward IXS: once the compression-induced heating has pushed $T$ well beyond the Debye temperature $\Theta_D$, the Stokes and anti-Stokes peak intensities reach parity, and no further useful information can be extracted from their ratio.

The third class of temperature-diagnostic techniques under active consideration involves measuring a sample's Debye-Waller factor directly from its x-ray diffraction pattern\cite{Murphy2008,Sharma2021}. The most conspicuous effect of temperature on the coherent scattering comes via the Bragg peaks, whose intensities decay exponentially with $M(T)$. However, this deceptively simple functional relationship between the signal and the temperature is difficult to exploit in reality owing to crystallographic texture. The intensity of each Bragg peak is dictated principally by the number of crystallites that meet the appropriate Bragg condition. That number will vary from shot-to-shot both stochastically (due to the x-ray beam sampling finitely many grains) and systematically (due to the plastic deformation effected by the compression\cite{Suggit2012,Wehrenberg2017,Sliwa2018}), making measurement of the Debye-Waller factor from the elastic scattering alone challenging for textured targets.

Only recently has it become apparent that this strong texture sensitivity can be circumvented by looking not \emph{at} the peaks, but \emph{between} them. As the intensity of the elastic scattering diminishes with increasing temperature, there is a concomitant increase in the thermal diffuse scattering (TDS), a largely structureless signal sitting beneath and between the Bragg peaks that is essentially the spectrally-integrated aforementioned IXS\cite{Hartley2021}. That TDS could serve as a texture-robust temperature diagnostic was not realized until a recent experiment undertaken by the present authors at EuXFEL\cite{Wark2025}. In that study, the authors obtained single-shot, femtosecond diffraction from rolled Cu foils shock-compressed to up to 140~GPa, and fitted the diffuse component of the signal to the classic model of TDS owed to Warren\cite{Warren1953}. The inferred temperatures were fully consistent with Hugoniot temperatures predicted by equations of state SESAME 3336 and LEOS 290 to within a statistical error comparable with that of EXAFS. Yet more remarkable was the observation that although Warren's TDS model was built to describe powders (i.e., polycrystals with no preferred crystallographic orientation), it predicted the form of the TDS remarkably well for textured foils, regardless of their compression state. These results raise the tantalizing prospect of a single-shot, texture-insensitive temperature diagnostic readily fieldable at XFEL facilities.

The paucity of current thermometric techniques makes it particularly important that we fully characterize this new TDS-based diagnostic, such that its limitations can be understood. To that end, we will present a new texture-aware model of thermal diffuse scattering for arbitrarily textured cubic polycrystals. Our main aims are to explain why Warren's model is so successful for textured samples, and to determine whether the TDS prediction can be made better yet by integrating texture into the model. We will then present a systematic study of how changes in crystallographic texture influence the form of the TDS signal (a study that, to our knowledge, has not been performed even in the context of `conventional' materials science), and thus explain the empirical observation that TDS is largely unaffected by the texture-related effects normally felt so acutely in dynamic-compression experiments. In so doing, we take the opportunity to give a full mathematical exposition of our thermal-diffuse-scattering model.

The main body of the paper is divided into two sections. In Sec.~\ref{sec:maths}, we first present the theory behind our texture-aware thermal-diffuse-scattering model in its entirety, explaining how we bridge the gap between single- and polycrystal scattering and how we incorporate higher-order scattering effects. In Sec.~\ref{sec:results}, we go on to explore the implications of our model as applied to dynamic-compression experiments. We begin by briefly presenting the x-ray scattering patterns and crystallographic texture of the rolled Cu foils used in the experiment detailed in Ref.~\onlinecite{Wark2025} in Secs.~\ref{sec:representative-data} and \ref{sec:starting-texture}, respectively. In Sec.~\ref{sec:powder-vs-rolled}, we then compare the TDS signals from perfectly random powders with those from polycrystals with a rolling texture. We then examine the stochastic changes to the TDS caused by finite grain-sampling in Sec.~\ref{sec:grain-statistics} and the systematic changes caused by plasticity in Sec.~\ref{sec:plasticity}. Finally, we discuss the implications of our study in Sec.~\ref{sec:discussion} before concluding in Sec.~\ref{sec:conclusion}.

\section{Thermal diffuse scattering model\label{sec:maths}}

\subsection{Scattering from a single crystal\label{sec:monocrystal-scattering}}
Before considering the role of crystallographic texture, we will first recap the origin of (first-order) thermal diffuse scattering from a perfect single crystal. The single-crystal scattering function is fundamentally the kernel over which we integrate to generate the scattering signal from a full polycrystalline aggregate. It is this function that encodes the detailed atom-atom correlations within each coherently scattering domain, and thus predicts the overall magnitude, structure, and temperature dependence of thermal diffuse scattering. Our goal is not to present a complete derivation -- which can be found in the reference texts of James~\cite{James1948} and Warren~\cite{Warren1990} -- but to highlight `milestones' in the model's construction and the simplifying assumptions underpinning them.

The elementary expression for the structure factor of a monatomic group of $N_a$ atoms with instantaneous positions $\{\mathbf{r}_n\}$ at reciprocal-space point $\mathbf{q}$ is
\begin{equation}
    \tilde{s}(\mathbf{q}) = f^2(\mathbf{q})\sum_{n=1}^{N_a} \left|e^{-i\mathbf{q}\cdot\mathbf{r}_n}\right|^2\ ,
\end{equation}
where $f(\mathbf{q})$ is the atomic form factor. We decompose the position of each atom into its equilibrium position $\mathbf{R}_n$ and its thermally induced displacement $\mathbf{u}_n$ before time-averaging and invoking the harmonic approximation, which yields
\begin{equation}\label{eq:structure_master}
    s(\mathbf{q}) = f^2(\mathbf{q})\sum_{n,m} e^{-i\mathbf{q}\cdot\mathbf{R}_{nm}} e^{-\frac{1}{2}\langle p_{nm}^2(\mathbf{q})\rangle}\ ,
\end{equation}
where $\mathbf{R}_{nm} = \mathbf{R}_n - \mathbf{R}_m$ and $p_{nm}(\mathbf{q}) = \mathbf{q}\cdot(\mathbf{u}_n - \mathbf{u}_m)$. The central problem of diffuse-scattering theory is to derive a closed-form expression for the thermal-displacement factor of $\exp(-\langle p_{nm}^2\rangle/2)$ that faithfully models long-range correlations between atomic displacements.

To this end, we populate the crystal with an equilibrium distribution of noninteracting phonons, each of which brings about a characteristic displacement field of
\begin{equation}
    \mathbf{u}_{\mathbf{g}j} = \mathbf{e}_{\mathbf{g}j}a_{\mathbf{g}j}\cos(\mathbf{g}\cdot\mathbf{r} - \omega_{\mathbf{g}j}t - \delta_{\mathbf{g}j})\ ,
\end{equation}
where $\mathbf{g}$ and $j$ denote the wavevector and polarization state of the phonon, $\mathbf{e}_{\mathbf{g}j}$ its polarization vector, $a_{\mathbf{g}j}$ its amplitude, $\omega_{\mathbf{g}j}$ its angular frequency, and $\delta_{\mathbf{g}j}$ a rapidly, randomly varying phase. If each atom's displacement is expressed as a linear combination of displacements from the entire phonon distribution (whose phases are assumed to be uncorrelated), one finds that
\begin{equation}
    \langle p_{nm}^2(\mathbf{q})\rangle = \sum_{\mathbf{g},j} (\mathbf{q}\cdot\mathbf{e}_{\mathbf{g}j})^2 a_{\mathbf{g}j}^2 \left[1 - \cos(\mathbf{g}\cdot\mathbf{R}_{nm})\right]\ .
\end{equation}
The amplitudes $\{a_{\mathbf{g}j}\}$ are in turn derived from the Bose-Einstein statistics governing the phonon populations. 
By equating the energies of a phonon mode predicted from its treatment as either a classical normal mode of oscillation or a quantum harmonic oscillator, we infer that the phonon amplitudes obey the equation
\begin{equation}
    a_{\mathbf{g}j}^2 = \frac{\hbar}{N_a m \omega_{\mathbf{g}j}}\coth\left(\frac{1}{2}\frac{\hbar\omega_{\mathbf{g}j}}{k_B T}\right)\ ,
\end{equation}
where $m$ is the atomic mass.

Following convention, the phonon amplitudes are re-parametrized via the coefficients
\begin{equation}
    G_{\mathbf{g}j}(\mathbf{q}) = \frac{1}{2}(\mathbf{q}\cdot\mathbf{e}_{\mathbf{g}j})^2a_{\mathbf{g}j}^2\ ,
\end{equation}
allowing us to express the thermal-displacement factor as
\begin{equation}
    e^{-\frac{1}{2}\langle p_{nm}^2(\mathbf{q})\rangle} = e^{-\sum_{\mathbf{g},j} G_{\mathbf{g}j}} e^{\sum_{\mathbf{g},j} G_{\mathbf{g}j} \cos(\mathbf{g}\cdot\mathbf{R}_{nm})}\ ,
\end{equation}
where the $\{G_{\mathbf{g}j}\}$ are closed-form functions encoding the phonon dispersion relation and population statistics. We note in passing that the mean-squared displacement of any given atom from its equilibrium position in the direction of $\mathbf{q}$ is given in the phonon formalism by
\begin{equation}
    \langle (\mathbf{q}\cdot\mathbf{u}_n)^2\rangle = \sum_{\mathbf{g},j} (\mathbf{q}\cdot\mathbf{e}_{\mathbf{g}j})^2 a_{\mathbf{g}j}^2\ ,
\end{equation}
whence it follows that
\begin{subequations}
\begin{align}
    e^{-\sum_{\mathbf{g},j} G_{\mathbf{g}j}} &= e^{-\frac{1}{2}\langle (\mathbf{q}\cdot\mathbf{u}_n)^2 \rangle} \\
    &\equiv e^{-2M(\mathbf{q})}\ ,
\end{align}
\end{subequations}
where $M(\mathbf{q})$ is the \textit{Debye-Waller factor}, the single most important parameter in the model, and the one that we ultimately measure experimentally.

\begin{figure}
    \centering
    \includegraphics{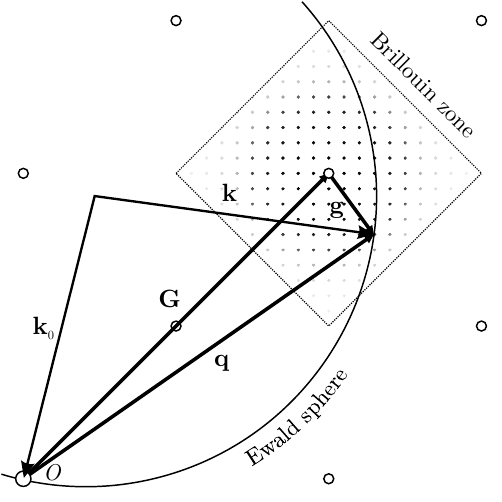}
    \caption{The first-order thermal diffuse scattering condition. At a given reciprocal lattice point $\mathbf{q} = \mathbf{k} - \mathbf{k}_0$ on the Ewald sphere, scattering is brought about with the aid of phonons in the vicinity of a reciprocal lattice vector $\mathbf{G}$ with wavevector $\mathbf{g}$ satisfying $\mathbf{q} - \mathbf{G} =\pm\mathbf{g}$. Hollow points indicate reciprocal lattice vectors. Filled points represent phonon wavevectors, the diminishing opacity of which indicates the falling phonon population further from the Brillouin zone center.}
    \label{fig:single-crystal}
\end{figure}

Incorporating these phonon-based functions into Eq.~(\ref{eq:structure_master}) (and henceforth dropping the argument notation for $f$, $M$, and $G_{\mathbf{g}j}$, all of which are understood to depend on $\mathbf{q}$) gives
\begin{equation}\label{eq:structure_preexpansion}
    s(\mathbf{q}) = f^2 e^{-2M}\sum_{n,m} e^{-i\mathbf{q}\cdot\mathbf{R}_{nm}}e^{\sum_{\mathbf{g},j} G_{\mathbf{g}j}\cos(\mathbf{g}\cdot\mathbf{R}_{nm})}\ .
\end{equation}
We now separate the single-crystal scattering signal into its elastic and thermal-diffuse components by expanding the second factor in the summand of Eq.~(\ref{eq:structure_preexpansion}) as an ascending power series (only the first two terms of which we will attempt to calculate explicitly):
\begin{widetext}
\begin{equation}
    s(\mathbf{q}) = f^2 e^{-2M}\sum_{n,m}e^{-i\mathbf{q}\cdot\mathbf{R}_{nm}}\Bigg[1 + \underbrace{\sum_{\mathbf{g},j}G_{\mathbf{g}j}\cos(\mathbf{g}\cdot\mathbf{R}_{nm})}_{\text{1st-order thermal diffuse}} + \underbrace{\frac{1}{2}\sum_{\mathbf{g},\mathbf{g}',j,j'}G_{\mathbf{g}j}G_{\mathbf{g}'j'}\cos(\mathbf{g}\cdot\mathbf{R}_{nm})\cos(\mathbf{g}'\cdot\mathbf{R}_{nm})}_{\text{2nd-order thermal diffuse}} +\ldots\Bigg]\ .
    \label{eq:big-expansion}
\end{equation}
\end{widetext}
The first contribution,
\begin{subequations}
\begin{align}
    s_0(\mathbf{q}) &= f^2 e^{-2M}\sum_{n,m}e^{-i\mathbf{q}\cdot\mathbf{R}_{nm}} \\
    &= f^2 e^{-2M} I_0(\mathbf{q})\ ,
\end{align}
\end{subequations}
expresses the elastic scattering, which takes the form of the equilibrium ionic structure factor $I_0(\mathbf{q})$ (a sharply localized function that is maximal at any reciprocal lattice vector $\mathbf{G}$) modulated by the atomic form factor and diminished by a factor of $e^{-2M}$. The second contribution,
\begin{equation}
    s_1(\mathbf{q}) = f^2e^{-2M}\sum_{n,m}e^{-i\mathbf{q}\cdot\mathbf{R}_{nm}}
    \sum_{\mathbf{g},j} G_{\mathbf{g}j}\cos(\mathbf{g}\cdot\mathbf{R}_{nm})\ ,
\end{equation}
constitutes the leading-order contribution to the thermal diffuse scattering. Its structure is made more transparent by rendering the cosine function in its complex-exponential form, giving
\begin{equation}\label{eq:first-order-general}
    s_1(\mathbf{q}) = f^2 e^{-2M}\frac{1}{2}\sum_{\mathbf{g},j}G_{\mathbf{g}j}\left[I_0(\mathbf{q}+\mathbf{g}) + I_0(\mathbf{q}-\mathbf{g})\right]\ .
\end{equation}
Hence, each phonon mode only contributes appreciably where its wavevector is such that $\mathbf{q}\pm\mathbf{g} = \mathbf{G}$, where $\mathbf{G}$ is any reciprocal lattice vector. It is thus instructive to picture every reciprocal lattice vector as being `dressed' by a Brillouin zone populated with phonons, as in Fig.~\ref{fig:single-crystal}; every scattering vector is effectively surrounded by a cloud of scattering intensity whose extent is that of the Brillouin zone and whose intensity grows with the phonon population at that point in the zone. We conclude that first-order scattering from phonons is therefore (1) diffuse (indeed, there is no point in reciprocal space that cannot be accessed from a scattering vector $\mathbf{G}$ with the assistance of a phonon wavevector $\mathbf{g}$) and (2) marginally structured (the intensity being maximal at the reciprocal lattice vectors themselves -- like the elastic scattering -- due to acoustic phonons being most populous).

To make implementation of Eq.~(\ref{eq:first-order-general}) more practicable, we appeal to Warren's model of first-order thermal diffuse scattering for cubic crystals~\cite{Warren1953}, which makes a number of simplifications to the Brillouin zone structure. First, the crystal temperature is taken to be sufficiently high (relative to its \textit{Debye temperature} $\Theta_D$) that all phonon population numbers tend toward their classical limit. Second, all phonons are modeled with an identical phase velocity. Third, the Brillouin zone is approximated as a sphere with such a radius $q_B$ as is required to give it the correct (inverse) volume for the given crystal structure; for a face-centered-cubic (fcc) crystal with cubic lattice constant $a$,
\begin{equation}\label{eq:brillouin-zone}
    q_B = \frac{2\pi}{a}\left(\frac{3}{\pi}\right)^{\frac{1}{3}}\ .
\end{equation}
Warren's final expression for the first-order diffuse scattering in the vicinity of reciprocal lattice vector $\mathbf{G}$ -- which we will simply quote -- reads
\begin{equation}\label{eq:first-order-specific}
    s_1(\mathbf{q}|\mathbf{G}) = N_a f^2 2Me^{-2M}W(\mathbf{q}-\mathbf{G})
\end{equation}
where the Debye-Waller factor takes its classical form (which is most accurate when $T>\Theta_D$):
\begin{equation}
    M(\mathbf{q}) = \frac{6h^2}{mk_B}\frac{T}{\Theta_D^2}\left(\frac{\sin\theta}{\lambda}\right)^2\ ,
\end{equation}
with $\theta$ being the Bragg angle corresponding to scattering vector $\mathbf{q}$, and where the \textit{Warren kernel} $W$ describing the distribution of diffuse scattering around each scattering vector is simply
\begin{equation}\label{eq:warren-kernel}
    W(\mathbf{k}) = \begin{cases}
    \frac{1}{3}\frac{q_B^2}{k^2} & \text{for}\quad 0<k\le q_B\ , \\
    0 & \text{for}\quad k>q_B\ .
    \end{cases}
\end{equation}
The salient feature of the first-order TDS is its temperature scaling via a factor of $2Me^{-2M}$. In the phonon-assisted--scattering picture, the $e^{-2M}$ term reflects the diminishing strength of the elastic scattering vectors at the center of each Brillouin zone, while the $2M$ term expresses the growing phonon population within each zone; generally, for hard x-rays scattered into the forward direction, the latter term dominates, and the first-order thermal diffuse scattering increases with temperature.

Finally, we note that to correctly predict the magnitude of the elastic scattering $s_0$ relative to the first-order diffuse scattering $s_1$, the ionic structure factor $I_0(\mathbf{q})$ must be normalized appropriately. If we express the elastic scattering distribution around a single scattering vector $\mathbf{G}$ as
\begin{equation}\label{eq:elastic-scattering}
    s_0(\mathbf{q}|\mathbf{G}) = N_a f^2 e^{-2M}J(\mathbf{q} - \mathbf{G})\ ,
\end{equation}
where $J$ is some plausible \textit{shape function} (essentially the three-dimensional analog of a lineshape), then $J$ must satisfy the constraint
\begin{equation}\label{eq:shape-function-integral}
    \int_{\mathbb{R}^3} d^3\mathbf{k}\,J(\mathbf{k}) = N_b\left(\frac{2\pi}{a}\right)^3\ ,
\end{equation}
where $N_b$ is the number of atoms in the conventional unit cell (4 for an fcc crystal).

To summarize, the elastic and first-order thermal diffuse scattering intensities at point $\mathbf{q}$ in the vicinity of a reciprocal lattice vector $\mathbf{G}$ take the form
\begin{equation}\label{eq:kernel}
    s(\mathbf{q}|\mathbf{G}) = \underbrace{N_a f^2 e^{-2M}J(\mathbf{q}-\mathbf{G})}_{s_0(\mathbf{q}|\mathbf{G})} + \underbrace{N_a f^2 2Me^{-2M}W(\mathbf{q}-\mathbf{G})}_{s_1(\mathbf{q}|\mathbf{G})}
\end{equation}
where $N_a$ is the number of atoms per grain, $f(\mathbf{q})$ is the atomic form factor, $M(\mathbf{q})$ is the Debye-Waller factor, and where the shape function $J$ and Warren kernel $W$ express the distribution of each type of scattering around any given reciprocal lattice vector; the elastic scattering $s_0$ is extremely concentrated, while the thermal diffuse scattering $s_1$ decays with a relatively slow inverse-square form. We now go on to discuss how we calculate the scattering intensity from a full polycrystal.

\subsection{Scattering from a polycrystal\label{sec:polycrystal-scattering}}
Calculating the total scattering from a polycrystalline aggregate is a matter of integrating contributions from every reciprocal lattice vector from every grain. In other words, one must know the full distribution of scattering vectors in reciprocal space.

\begin{figure*}
    \centering
    \includegraphics{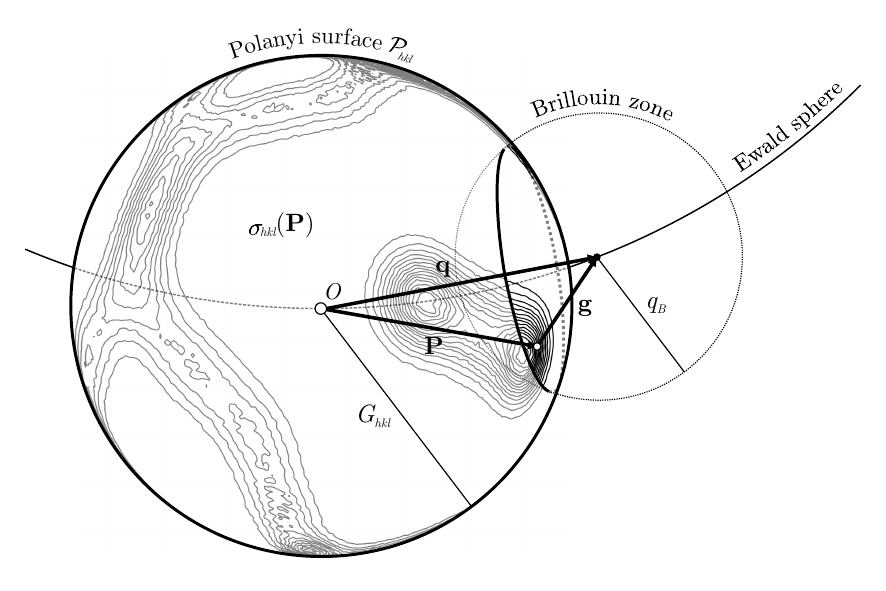}
    \caption{First-order thermal diffuse scattering calculation for a polycrystal. Largest sphere depicts Polanyi surface $\mathcal{P}_{hkl}$ accommodating an anisotropic distribution of scattering vectors from the $\{hkl\}$ family of planes, with $\sigma_{hkl}(\mathbf{P})$ scattering vectors per unit solid angle at point $\mathbf{P}$. First-order scattering at point $\mathbf{q}$ on the Ewald sphere is calculated by integrating the elementary scattering function $s_1(\mathbf{q}|\mathbf{P})$ from Eq.~(\ref{eq:first-order-specific}) over the locus of scattering vectors on $\mathcal{P}_{hkl}$ falling within a spherical Brillouin zone of radius $q_B$ centered on $\mathbf{q}$. Loci are shown to scale for the $\{111\}$ Polanyi surface ($G_{111}=3.01$~\AA\textsuperscript{$-1$}) of a face-centered-cubic polycrystal with lattice constant $a=3.615$~\AA\ ($q_B=1.71$~\AA\textsuperscript{$-1$}) probed by an 18~keV x-ray source ($E/\hbar c=9.12$~\AA\textsuperscript{$-1$}).}
    \label{fig:poly-crystal}
\end{figure*}

Under ambient conditions, the reciprocal lattice vectors of a cubic polycrystal reside on a set of concentric spheres called \textit{Polanyi surfaces}. Each Polanyi surface $\mathcal{P}_{hkl}$ accommodates scattering vectors belonging to a particular family of planes denoted by the Miller indices $\{hkl\}$. These vectors have an associated magnitude $G_{hkl}$ (the radius of $\mathcal{P}_{hkl}$) and multiplicity $j_{hkl}$. We choose to describe the distribution of scattering vectors over each Polanyi surface via the number of scattering vectors per unit solid angle, $\sigma_{hkl}$, which must satisfy
\begin{equation}
    \iint\limits_{\mathcal{P}_{hkl}}d\Omega\,\sigma_{hkl} = N_g j_{hkl}\ ,
\end{equation}
where $N_g$ is the total number of grains in the polycrystal, which we assume throughout to be of identical size. Under hydrostatic conditions, all Polanyi densities $\{\sigma_{hkl}\}$ are derivable from a single orientation distribution function (ODF). The ODF generally depends on both the manner of preparation of the sample and the loading path to which it is subjected in the experiment.

We digress briefly to note that during dynamic uniaxial compression -- the scenario in which we ultimately wish to apply TDS as a temperature diagnostic -- there is no guarantee that the loci described by the scattering vectors remain spherical. In the simplest picture of compression, the unit cell contracts equally in all directions while the crystal structure is perfectly preserved. In reciprocal space, such hydrostatic compression manifests as an isotropic enlargement centered on the origin; the Polanyi surfaces expand, but remain spherical. Hence, a single ODF is sufficient to uniquely describe the polycrystal's crystallographic texture.

In reality, however, dynamic compression is often realized via a uniaxial loading path. Grains immediately behind the traveling compression front find themselves in a state of extreme deviatoric elastic strain, in which the interatomic spacing has decreased along the compression direction, but is unchanged in the transverse directions. Each grain will rapidly (over a few picoseconds\cite{Rudd2012}) relax towards a hydrostatic strain state via one of any number of shear-stress--relieving mechanisms (crystallographic slip\cite{Milathianaki2013}, deformation twinning\cite{Wehrenberg2017}, shear banding\cite{Cawkwell2008}, grain-grain interactions\cite{Heighway2019b}), but will inevitably stop somewhere short of a perfectly isotropic strain state, because the residual shear stress falls below the threshold needed to activate further deformation (an effect referred to in the shorthand as `strength').

One should therefore expect that the typical unit cell will remain marginally more compressed along the loading direction than along the transverse directions. As a result, the Polanyi surfaces will be distended into broadly ellipsoidal shapes, whose longest axis lies parallel to the compression direction; indeed, modeling the Polanyi surfaces as perfect ellipsoids \cite{Higginbotham2014} is a common approach to inferring the strength of shock-compressed metals\cite{Hawreliak2011,Foster2017,Foster2021}. However, even this model does not capture the true complexity of the post-shock scattering-vector distribution, which is not even guaranteed to form continuous surfaces. In this regime, an ODF alone is insufficient to characterize the polycrystal's reciprocal space, necessitating a higher-dimensional function that encodes the distribution of the local lattice spacings with orientation.

To keep the present model tractable, however, we will neglect strength effects, and assume that the scattering vectors continue to inhabit a set of concentric spherical surfaces, regardless of compression state. We do so principally to make better contact with the original model of Warren\cite{Warren1953}, which is predicated on every grain having a perfect cubic crystal structure. Our model is therefore notionally unsuitable for materials that exhibit `significant' strength upon compression (e.g., diamond\cite{McWilliams2010}).

Assuming the Polanyi surfaces remain spherical and that all grains scatter independently, the signal from the polycrystal may be expressed by the integral
\begin{equation}\label{eq:texture-integral}
    S(\mathbf{q}) = \sum_{hkl} \iint\limits_{\mathcal{P}_{hkl}} d\Omega\,\sigma_{hkl}(\mathbf{P})s(\mathbf{q}|\mathbf{P})\ .
\end{equation}
That is to say, each bundle of scattering vectors concentrated at point $\mathbf{P}$ on Polanyi surface $\mathcal{P}_{hkl}$ [of abundance $d\Omega\,\sigma_{hkl}(\mathbf{P})$] contributes $s(\mathbf{q}|\mathbf{P})$ per vector to the scattering pattern, where the functional form of $s$ is given by Eq.~(\ref{eq:kernel}). The structure of these integrals is practically identical for the elastic and first-order thermal diffuse contributions to $S(\mathbf{q})$ (which are additive), the only difference being whether $J$ or $W$ appears in the integrand.

While the integral is formally conducted over the entire Polanyi surface, functionally, the variable of integration $\mathbf{P}$ is restricted to those regions of the Polanyi surface close enough to reciprocal-space point $\mathbf{q}$ to contribute appreciably to the pattern. We show in Fig.~\ref{fig:poly-crystal} a schematic of the integral required to evaluate the first-order TDS at point $\mathbf{q}$ from a single Polanyi surface $\mathcal{P}_{hkl}$, for which we need only consider points within the Brillouin zone radius $q_B$ of $\mathbf{q}$, since the Warren kernel $W$ vanishes beyond that separation. Calculation of the elastic scattering is yet more localized, requiring contributions only from a narrow belt of scattering vectors situated in the vicinity of the intersection between the Polanyi and Ewald spheres (albeit with a greater resolution).

Borrowing terminology from Paskin\cite{Paskin1958,Chipman1959a}, we render the first-order TDS $S_1(\mathbf{q})$ in the form
\begin{subequations}\label{eq:paskin-factor}
\begin{align}
    S_1(\mathbf{q}) &= \sum_{hkl}\iint\limits_{\mathcal{P}_{hkl}}d\Omega\,\sigma_{hkl}(\mathbf{P})s_1(\mathbf{q}|\mathbf{P}) \\
    &= N_aN_gf^22Me^{-2M}\underbrace{\sum_{hkl}\iint\limits_{\mathcal{P}_{hkl}}d\Omega\,\hat{\sigma}_{hkl}(\mathbf{P})W(\mathbf{q}-\mathbf{P})}_{C_1(\mathbf{q})}\label{eq:thermal-textural_decomp}
\end{align}
\end{subequations}
where $C_1(\mathbf{q})$ is an intensive, temperature-independent, geometric factor that encapsulates the polycrystal's crystallographic texture (via the factor $\hat{\sigma}_{hkl} = \sigma_{hkl}/N_g$) and explains the characteristic structure of first-order TDS. We have validated the integration scheme outlined by Eq.~(\ref{eq:texture-integral}) by checking that it correctly reproduces the known, closed-formed geometric coefficient $C_1$ in select high-symmetry scenarios, including the case of a random powder originally treated by Warren, for which
\begin{equation}
    C_1(\mathbf{q}) = \sum_{hkl} \frac{1}{6}j_{hkl}\frac{q_B^2}{qG_{hkl}}\ln\left(\frac{q_B}{|q-G_{hkl}|}\right)\ .
    \label{eq:warren_paskin}
\end{equation}
Details of the integration validation are given in the Supplementary Material.

\subsection{Higher-order scattering}
We have thus far considered only first-order TDS. According to the expansion of $s(\mathbf{q})$ in Eq.~(\ref{eq:big-expansion}), there exist higher-order, diminishing contributions to the scattering that correspond physically to multi-phonon interactions with the incoming photon. In analogy with Eq.~(\ref{eq:paskin-factor}), the total thermal diffuse scattering can be expressed as a sum of $\ell$-phonon scattering events, which takes the form of a power series in $2M$:
\begin{subequations}
\begin{align}
    S_{\text{TD}}(\mathbf{q}) &= \sum_{\ell = 1}^\infty S_\ell(\mathbf{q}) \\
    &= N f^2 e^{-2M}\sum_{\ell=1}^\infty \frac{(2M)^\ell}{\ell!}C_\ell(\mathbf{q})\label{eq:TDS-power-series}
\end{align}
\end{subequations}
where $N=N_aN_g$ is the total number of scattering atoms and where $C_\ell$ encodes the structure of the $\ell$\textsuperscript{th}-order thermal diffuse scattering.

It is indeed possible to calculate the coefficients $\{C_\ell\}$ within the framework of the Debye model used here, but even the lowest-order terms have excessively complicated analytic form. To keep application of the model tractable, we use an accurate approximation owed to Borie\cite{Borie1961}, who observed that for an fcc powder,
\begin{equation}
    C_\ell \approx \begin{cases}
    \frac{1}{2}(1+C_1) & \text{for}\quad \ell=2\ , \\
    1 & \text{for}\quad \ell > 2\ .
    \end{cases} \label{eq:borie}
\end{equation}
where $C_1$ [defined in Eq.~(\ref{eq:paskin-factor})] describes the reciprocal-space structure of $S_1$. In other words, the second-order TDS is structured in much the same way as the first-order (albeit to a lesser extent), while all higher-order diffuse scattering signals are essentially structureless.

\begin{figure*}
    \includegraphics{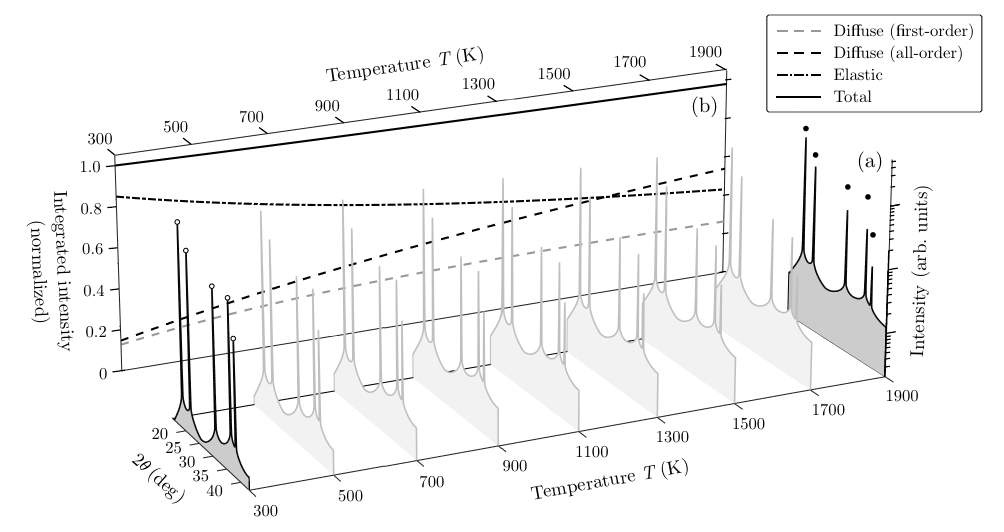}
    \caption{Conservation of total integrated scattering intensity during (isochoric) temperature increase. (a) Modeled diffraction from a powderlike Cu polycrystal illuminated by an 18~keV x-ray source as a function of temperature, including elastic and thermal-diffuse components (logarithmic scale). The intensities of the first five Bragg peaks at 300~K are shown by black data points above the same peaks at 1900~K for comparison. (b) Integrated diffraction intensities from the thermal diffuse and elastic components of scattering over the interval $2\theta\in[17,60]^\circ$ as a function of temperature (linear scale). Integrated intensities are normalized to the total (elastic + all-order thermal diffuse) integrated intensity at 300~K.}
    \label{fig:conservation}
\end{figure*}

Substituting Borie's structure coefficients into Eq.~(\ref{eq:TDS-power-series}) yields the approximation for the all-order diffuse scattering that we use throughout:
\begin{equation}
\begin{split}
    S_{\text{TD}}(\mathbf{q}) &= Nf^2(1-e^{-2M}) \\
    &+Nf^2e^{-2M}(2M+M^2)(C_1-1)\ .
\end{split}
\end{equation}
To justify Borie's approximation, we have validated that it satisfies Plancherel's theorem, which requires that the total scattering from a set of electrons -- integrated over all reciprocal space -- must be constant, regardless of their arrangement. This is to say that if we raise the temperature of the polycrystal, the attendant decrease in elastic scattering (owed to the enhanced Debye-Waller factor $2M$) must be compensated for exactly by an increase in thermal diffuse scattering. Strictly speaking, the conservation is exact only when the entire three-dimensional reciprocal space is integrated over, but the scattering over the Ewald sphere embedded in that space should still be strongly conservative. By showing that Plancherel's theorem is satisfied by our model, we demonstrate that both the temperature scaling of the scattering components and their relative magnitudes is correct.

In Fig.~\ref{fig:conservation}, we show a set of diffraction patterns from a powderlike Cu polycrystal simulated at temperatures between 300~K and 1900~K. The intensities are plotted on a logarithmic scale to expose the relatively weak TDS, whose maximum intensity at ambient temperatures is at least two orders of magnitude less than that of the Bragg peaks. We show in that same figure the total scattering intensities from the elastic and thermal-diffuse components (both at first order and at all orders) integrated over the scattering angle range $2\theta\in[17,60]^\circ$. As required, the diminishing elastic scattering intensity is balanced almost exactly by the growing thermal-diffuse background, such that the total scattering intensity is conserved to within 2\%. Note that, with the present parameters, the first-order diffuse scattering actually saturates and eventually diminishes beyond $\sim1800$~K, meaning the higher-order contributions are essential to maintain intensity conservation.

We will assume that Borie's approximation for the higher-order scattering is reasonable not only for powders, but also for moderately textured polycrystals. We will justify this claim \textit{a posteriori} in Sec.~\ref{sec:powder-vs-rolled}) where we compare our model predictions with experimental data.

\begin{figure*}
    \includegraphics{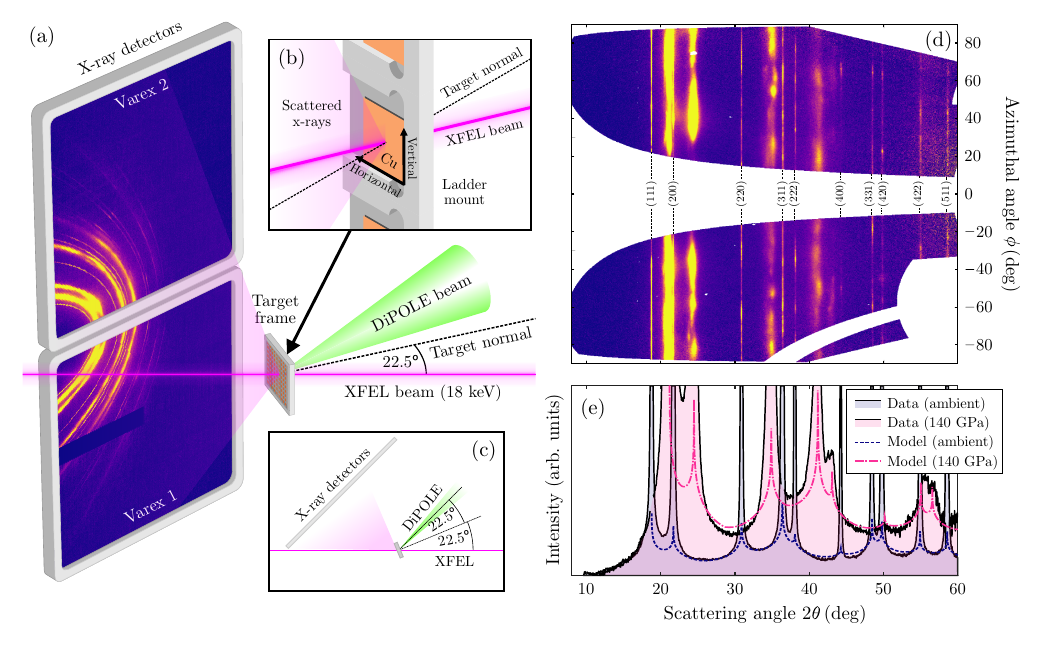}
    \caption{X-ray thermal diffuse scattering (TDS) patterns obtained from shock-compressed Cu at the High Energy Density (HED) instrument of the European X-Ray Free-Electron Laser (EuXFEL) \cite{Wark2025}. (a) Experimental configuration. Copper foils are dynamically loaded using the DiPOLE 100-X laser and simultaneously probed by an 18~keV photon beam inclined at $22.5^\circ$ to the target normal. Diffraction is captured on a pair of symmetrically displaced megapixel Varex detectors in transmission geometry. (b) Close-up of an individual Cu target viewed downstream, indicating the sense of the `vertical' and `horizontal' directions. (c) Configuration viewed in the horizontal plane. (d) Full diffraction pattern from Cu shock-compressed to 140~GPa, presented in $(2\theta,\phi)$-space. White regions either fall outside detector coverage, were obscured by shadows, or failed to have data recorded. (e) Azimuthally averaged scattering from an ambient (blue) and a driven (pink) shot shown at the scale of the TDS. Data are normalized to the incident photon flux, corrected to exclude polarization, per-pixel solid-angle, and all attenuation factors, and have had Compton scattering and scattering from the drive-side ablator layer subtracted. The inter-Bragg regions of each pattern are fitted to Warren's TDS model for a perfectly random power (dashed lines).}
    \label{fig:data-summary}
\end{figure*}

\section{Texture-sensitivity analysis\label{sec:results}}
Having constructed a texture-aware thermal-diffuse scattering model, we now aim to address the question of whether crystallographic texture significantly complicates the process of extracting the Debye-Waller factor (and hence temperature) from TDS measurements. To this end, we will address three related questions:
\begin{enumerate}
    \item Does the thermal diffuse scattering from a `typical', commercial rolled foil differ from that of an ideal random powder?
    \item Do statistical errors resulting from the x-ray beam sampling finitely many grains change the thermal diffuse scattering pattern?
    \item Do systematic texture changes caused by shock-induced plastic deformation change the thermal diffuse scattering?
\end{enumerate}
The backdrop against which we perform this investigation is the dynamic-compression study of Wark \textit{et al.}\cite{Wark2025}, in which TDS-based temperature measurements of shock-loaded rolled Cu foils were obtained from femtosecond x-ray diffraction patterns. Throughout, our model parameters will be chosen to match those of the experiment, the essential details of which we will recap in Sec.~\ref{sec:representative-data}. We will then describe the starting texture used for all of our simulations in Sec.~\ref{sec:starting-texture} before answering the questions above in Secs.~\ref{sec:powder-vs-rolled}-\ref{sec:plasticity}.

\subsection{Representative thermal diffuse scattering data\label{sec:representative-data}}
A complete account of the thermal-diffuse-scattering-based temperature measurements of shock-compressed Cu obtained at the inaugural DiPOLE 100-X experiment may be found in Ref.~\onlinecite{Wark2025} (along with accounts of parallel studies undertaken on Sn and C in Refs.~\onlinecite{Gorman2024} and \onlinecite{Kraus2025}, respectively); here, we recap only the essential details. The experiment was undertaken at the High Energy Density (HED) instrument of the European X-Ray Free-Electron Laser (EuXFEL) using the setup shown in Fig.~\ref{fig:data-summary}(a). Targets comprising a 50-$\upmu$m-thick Kapton-B (DuPont) ablator layer glued to 25~$\upmu$m of rolled Cu (Goodfellow) were shock-compressed up to pressures approaching 200~GPa using 10-ns pulses of frequency-doubled (515~nm) radiation from the DiPOLE laser containing up to 40~J. Shortly before shock breakout, the targets were probed with a 50~fs pulse of 18~keV x-rays traveling at $22.5^\circ$ to the target normal. Femtosecond diffraction patterns were recorded on a pair of Varex detectors situated downstream of the XFEL beam.

We show representative diffraction data from a single shot at close to 140~GPa shock pressure in Fig.~\ref{fig:data-summary}(d). That the Cu foil possesses a non-trivial crystallographic texture is clear from the azimuthal modulations in scattering intensity around the Debye-Scherrer rings. However, when we azimuthally average our diffraction patterns and examine the result at an intensity level approximately two orders of magnitude lower than that of the Bragg peaks [Fig.~\ref{fig:data-summary}(e)], we observe a clear thermal diffuse scattering signal whose structure is reproduced reasonably well by Warren's powder model. The accuracy of the prediction degrades at higher compression, an observation likely attributable -- at least in part -- to the growing effects of anharmonicity at higher shock temperatures (at least 3000~K at 140~GPa) not accounted for in our model. Regardless, Warren's powder model is perhaps surprisingly successful given the degree of texturing evident from the azimuthal structure of the Debye-Scherrer rings. Our goal is to explain this observation quantitatively by forward-modeling the TDS pattern expected of our textured specimens, which requires that we first understand their starting texture.

\subsection{Calculation of the starting texture\label{sec:starting-texture}}

\begin{figure}
    \includegraphics{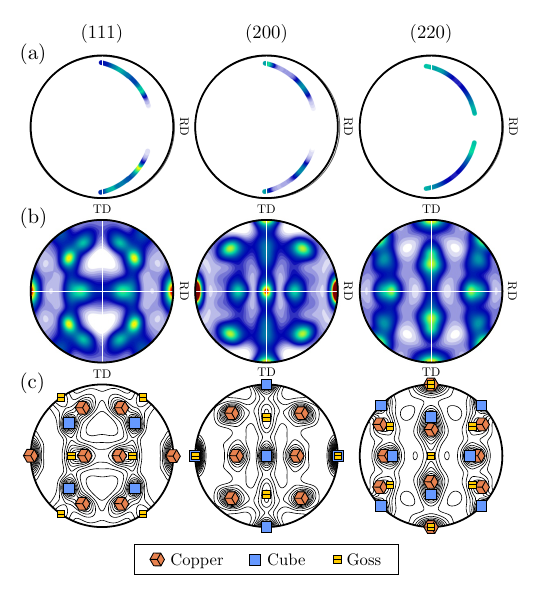}
    \caption{Starting texture of rolled Cu foils in the sample frame. (a) Equal-area, partial pole figures for the (111), (200), and (220) planes obtained by illuminating the foils with a quasimonochromatic 18~keV photon beam at $22.5^\circ$ to the target normal, shown using an equal-area projection. (b) Complete pole figures reconstructed using \textsc{mtex} \cite{MTEX}. (c) Contour-style pole figures with the high-symmetry copper, cube, and Goss orientations overlaid (along with their orthorhombic-symmetric equivalents). Labels RD and TD denote the inferred rolling and transverse directions, respectively.}
    \label{fig:starting-texture}
\end{figure}

To calculate the Cu targets' initial ODF (whence we can derive all Polanyi densities $\{\sigma_{hk}\}$), we first aggregated diffraction patterns from ten separate foils with the same mounting orientation on the target frame and took their average after correcting for extraneous factors such as polarization and attenuation. We then took each Debye-Scherrer ring in turn, and calculated the variation of the integrated intensity under the peak as a function of azimuthal angle $\phi$. Combining Eqs.~(\ref{eq:elastic-scattering}) and Eqs.~(\ref{eq:texture-integral}), one may show that if the shape function $J$ is sufficiently localized, the radially integrated intensity should vary according to the equation
\begin{equation}
    I_{hkl} = \left[N_a\left(\frac{\lambda}{a}\right)^2f^2e^{-2M}\frac{1}{\sin^2\theta_{hkl}\cos\theta_{hkl}}\right]\sigma_{hkl}\ .\label{eq:diffraction-to-texture}
\end{equation}
We can thus calculate the local density of scattering vectors $\sigma_{hkl}$ along the intersection between the Ewald and Polanyi spheres directly from the peak intensity $I_{hkl}$. Repeating this calculation for Debye-Scherrer rings with Miller indices between (111) and (420) inclusive yields eight partial pole figures evaluated in the x-ray frame. We then used the \textsc{mtex} software package \cite{MTEX} to estimate the single underlying ODF consistent with all eight pole figures. To expedite ODF reconstruction, we first actively rotated the partial pole figures by $22.5^\circ$ about the vertical direction to bring them into the sample frame, allowing us to exploit the global orthorhombic $(mmm)$ symmetry imparted to the foils by the rolling process.

We show in Figs.~\ref{fig:starting-texture}(a) and \ref{fig:starting-texture}(b) the partial and reconstructed pole figures, respectively, for the (111), (200), and (222) scattering planes. In the parlance of Sec.~\ref{sec:polycrystal-scattering}, the pole figure for the $(hkl)$ family of planes depicts (up to a multiplicative constant) the Polanyi density $\sigma_{hkl}$ over the upper hemisphere of Polanyi surface $\mathcal{P}_{hkl}$. We observe a starting texture that is broadly consistent with the $\beta$-fiber expected of rolled Cu\cite{Dillamore1964,Sidor2013,Kestens2016}. In Fig.~\ref{fig:starting-texture}(c), we show contour-style pole figures overlaid with the plane directions for particular high-symmetry grain orientations. We find that the ODF is essentially trimodal, with the three dominant orientations being the copper [expressed with Euler angles $(\varphi_1,\Phi,\varphi_2) = (90,35,45)^\circ$] and Goss [$(0,45,90)^\circ$] orientations -- both of which constitute part of the $\beta$-fiber -- in addition to the standard cube [$(0,0,0)^\circ$]. For the set of targets used to reconstruct the ODF, we infer that the foil rolling direction (RD) was horizontally oriented, with the sense of `horizontal' being shown in Fig.~\ref{fig:data-summary}(b). We assume that every target can reasonably be assumed to have the same starting texture on the basis that each one was cut from the same sheet of Cu, allowing us to use a single master ODF in all of the modeling that follows.

\subsection{Comparison between powder and rolling-texture signals\label{sec:powder-vs-rolled}}

\begin{figure*}
    \includegraphics{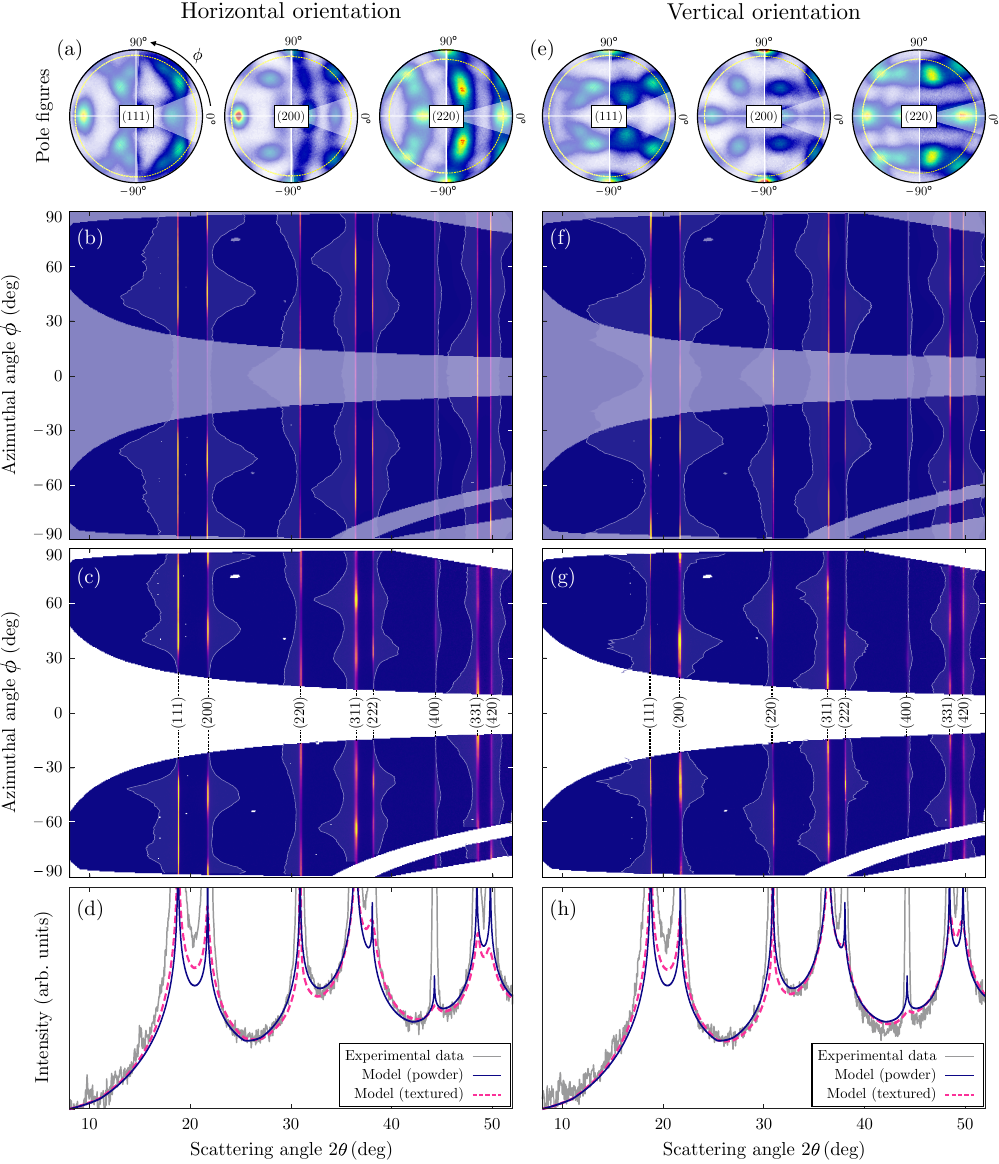}
    \caption{Modeling of thermal diffuse scattering (TDS) from rolled Cu foils under ambient conditions. (a) Equal-area pole figures showing the Polanyi density $\sigma_{hkl}$ of the (111), (200), and (220) planes in the x-ray frame for targets with their rolling direction oriented horizontally. Dashed yellow lines show the loci of planes that meet the Bragg condition (where the Ewald sphere and Polanyi surface intersect). Azimuthal regions not covered by the detectors are covered by a semitransparent white mask. (b) Modeled azimuthally resolved diffraction pattern in $(2\theta,\phi)$ space. Azimuthal intensity variations around each Debye-Scherrer ring are indicated by overlaid lineouts. Data has been divided by atomic-form and Debye-Waller factors to make higher-order peaks more easily visible. (c) Experimental azimuthally resolved diffraction pattern. (d) Inter-peak, azimuthally averaged experimental diffraction (gray) compared with modeled thermal diffuse scattering (TDS) for a perfectly random powder (dark blue) and for the textured polycrystal (pink, dashed). (e-h) As (a-d), but for targets with a vertical rolling direction.}
    \label{fig:master}
\end{figure*}

We provide in Fig.~\ref{fig:master} an overview of the comparison between the x-ray scattering measured from uncompressed Cu foils in the experiment described in Sec.~\ref{sec:representative-data} with the predictions of our texture-aware diffraction model. To enrich the comparison, we show results for sets of targets with two different mounting orientations, such that their rolling direction (RD) is either oriented horizontally or vertically, defined in the sense shown in Fig.~\ref{fig:data-summary}(b). By probing targets with inequivalent orientations with respect to the incoming x-ray beam, we can test our model's accuracy over a greater volume of reciprocal space. We first consider the targets with a horizontal RD from which the starting texture was calculated.

The elastic scattering from the horizontally oriented targets may be understood with reference to Fig.~\ref{fig:master}(a), which shows low-index pole figures in the x-ray frame. Overlaid on these figures are the one-dimensional loci of intersection between the Ewald and Polanyi spheres where the Bragg condition is met; at the azimuthal scattering angles where these loci traverse a region of the Polanyi surface densely populated with scattering vectors, we expect to see strong elastic scattering. The full diffraction pattern in $(2\theta,\phi)$-space is shown in Fig.~\ref{fig:master}(b), with lineouts showing the azimuthal intensity variation around each Debye-Scherrer ring superimposed. The predicted pattern may be compared with the average of ten experimental diffraction patterns from ambient Cu shown in Fig.~\ref{fig:master}(c). In both patterns, we have divided by the atomic-form and Debye-Waller factors that would otherwise attenuate the scattering with increasing $2\theta$ to better expose the structure of high-index Bragg peaks. On the whole, we capture the observed azimuthal structure of the Debye-Scherrer rings successfully.

With the elastic scattering prediction validated, we turn to the thermal diffuse scattering. In Fig.~\ref{fig:master}(d), we plot the azimuthally averaged experimental diffraction at the scale of the TDS alongside two predicted signals, the first assuming a perfectly random powder, the second factoring in the full crystallographic texture; the two modeled TDS signals are normalized such that the total number of scattering grains $N_g$ is equal for both. Several conclusions may be drawn from this plot: first, both modeled TDS signals reproduce the experimentally measured intensity very well; second, the predicted TDS is remarkably similar for the textured and untextured polycrystals; third, while the differences between the two are slight, it is in fact possible to discern that the TDS model incorporating the target texture provides a better fit than does the powder model. Examining the TDS intensity in the intervals between the (220)/(311), (222)/(400), and (400)/(331) peaks, we see that the powder model \mbox{over-,} \mbox{under-,} and over-predicts the measured TDS, respectively, while the texture-aware model provides a stronger match to the data in all three intervals.

We repeat all of the foregoing analysis on samples whose RD is instead aligned vertically in Figs.~\ref{fig:master}(e-h). The ODF of these samples is not fitted explicitly, but is rather obtained simply by rotating the ODF of the horizontally oriented foils by $90^\circ$ about the normal direction (ND). We see in Figs.~\ref{fig:master}(f) and \ref{fig:master}(g) that the overall structure of the diffraction pattern is reproduced reasonably well by our model, though with less accuracy than for the horizontally oriented specimens, which we attribute to our being unable to reconstruct the ODF perfectly from a single `still image'. Comparing the modeled and measured TDS signals in Fig.~\ref{fig:master}(h), we observe once again that while the fidelity of the texture-aware TDS model is greater, its predictions differ from those of the powder model only marginally. Indeed, by comparing Figs.~\ref{fig:master}(d) and \ref{fig:master}(h), we see that the thermal diffuse scattering patterns from the horizontally and vertically oriented foils are similar not only to that of an ideal powder, but also to one another. We conclude that the structure of the azimuthally integrated TDS is -- for these moderately textured rolled foils, at least -- remarkably insensitive to the details of the polycrystal's orientation distribution.

\begin{figure}
    \centering
    \includegraphics{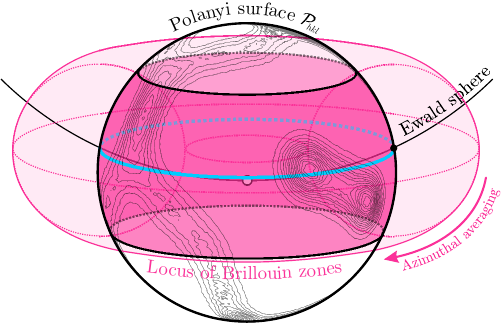}
    \caption{Schematic illustrating the belt of scattering vectors sampled via measurement of azimuthally averaged scattering. Blue curve indicates the quasi-1D locus sampled by elastic scattering, found by calculating the intersection between the Ewald sphere and the Polanyi surfaces. Pink band indicates the 2D locus sampled by thermal diffuse scattering (at $q$-values that also meet the Bragg condition), found by calculating the overlap of an extruded torus of Brillouin zones and the Ewald sphere. Loci are shown to scale for the $\{111\}$ Polanyi surface of ambient Cu probed by an 18~keV x-ray beam, for which at most 54\% of $\mathcal{P}_{111}$ is sampled.}
    \label{fig:torus}
\end{figure}

The intuitive explanation for this insensitivity can be understood with reference to Fig.~\ref{fig:torus}. The (first-order) TDS at a single point $\mathbf{q}$ on the Ewald sphere [i.e., a single $(2\theta,\phi)$] is given by a weighted integral of the scattering-vector density $\sigma_{hkl}$ over those regions of the Polanyi surfaces within a Brillouin-zone radius $q_B$ of $\mathbf{q}$. The total fraction of the Polanyi surface that these regions constitute is fairly small; for example, the spherical cap of participating (111) scattering vectors for the experimental setup considered here (shown in Fig.~\ref{fig:poly-crystal}) constitutes at most 8\% of $\mathcal{P}_{111}$. However, when we azimuthally average the TDS, that fraction grows considerably. By integrating the intensity at all $\mathbf{q}$-points around the Ewald sphere sharing the same $2\theta$, that spherical cap of participating scattering vectors is smeared into the ‘belt’ shown in Fig.~\ref{fig:torus}; for the (111) vectors, that belt constitutes as much as 54\% of the total Polanyi surface. Hence, the total proportion of the Polanyi surface that participates in thermal diffuse scattering is large.

Now suppose we started with a perfectly random powder, and then gradually imparted to it some degree of texture. In reciprocal space, the introduction of texture amounts to the conservative redistribution of scattering vectors around the Polanyi surfaces.  If the belt of scattering vectors participating in TDS is sufficiently broad, the total number of vectors that migrate out of the belt due to the texture change should be approximately equal to the number that migrate in; the bigger the belt, the better is this compensatory effect. Note also that if a change in texture diminishes the TDS measurable from one Polanyi surface, one usually observes a commensurate increase in scattering from a different Polanyi surface; since the TDS at almost every $\mathbf{q}$-point includes contributions from multiple Polanyi surfaces, this introduces another compensatory effect that further reduces the sensitivity of the TDS to texture (away from its maxima, at least). It is for these reasons that the TDS patterns from our rolled Cu foils obtained at the HED instrument differ only marginally from that of a perfect powder.

\begin{figure}
    \includegraphics{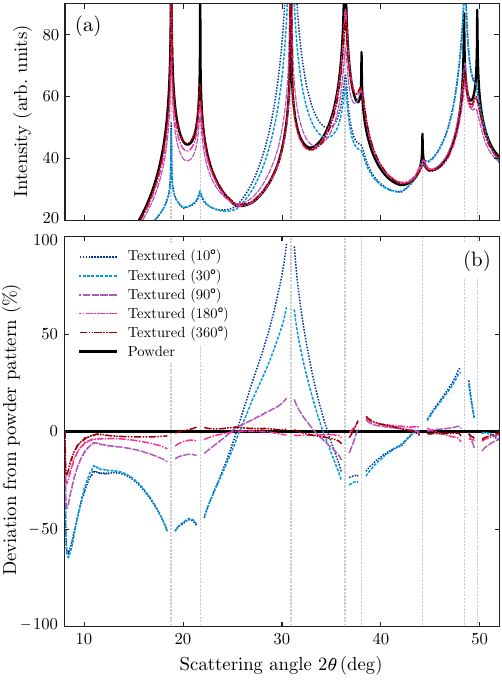}
    \caption{Dependence of the modeled azimuthally averaged thermal diffuse scattering signal from a horizontally oriented Cu foil on the size of the azimuthal integration window, varying between $\phi\in[-5,5]^\circ$ and $\phi\in[-180,180]^\circ$ (i.e., full $360^\circ$ coverage). (a) Absolute scattering signals compared with the signal from a powder containing the same number of grains. (b) Fractional difference between the azimuthally averaged signals and the powder signal. Regions in the immediate vicinity of the maxima where differences are largely attributable to numerical-integration artifacts are masked out.}
    \label{fig:patio-doors}
\end{figure}

\begin{figure}
    \centering
    \includegraphics{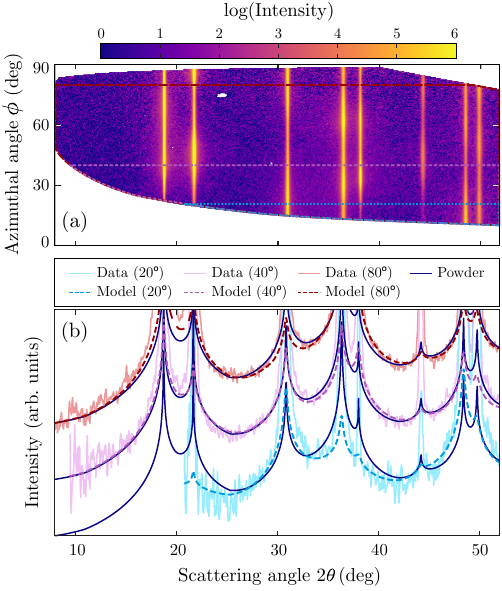}
    \caption{Dependence of the experimental azimuthally averaged thermal diffuse scattering (TDS) signal from a horizontally oriented Cu foil on the size of the integration window, spanning $\phi\in[0,20]^\circ$, $[0,40]^\circ$, or $[0,80]^\circ$. (a) Experimental azimuthally resolved diffraction pattern in $(2\theta,\phi)$ on the upper detector only, with intensity plotted on a logarithmic scale. Colored lines indicate the blade-shaped azimuthal averaging regions. (b) Experimental and modeled azimuthally averaged TDS signals for each integration region compared with the signal from a perfect powder. TDS patterns from each integration region are offset vertically for clarity.}
    \label{fig:expatio-doors}
\end{figure}

Key to this argument is the size of the azimuthal range over which the TDS is averaged, or, in other words, how far the belt of detectable participating scattering vectors wraps around the Polanyi surfaces. The standard detector geometry at HED offers reasonably generous angular coverage; as shown in Fig.~\ref{fig:data-summary}, each of the two Varex detector covers an azimuthal domain of approximately $60^\circ$ at every $2\theta$. Were this azimuthal range considerably reduced, we should \emph{not} expect the TDS pattern to resemble that of a powder. To demonstrate this, we can model how the azimuthally averaged TDS from our (horizontally oriented) rolled Cu foils varies with the angular range over which the averaging is performed. For convenience, we choose to average over the symmetric domain $\phi\in[-\Delta\phi/2,\Delta\phi/2]$, sampling domains sizes of $\Delta\phi=10^\circ,30^\circ,90^\circ,180,^\circ$, and $360^\circ$. The predicted patterns are show in Fig.~\ref{fig:patio-doors}(a), along with the ideal powder signal for comparison. We also plot in Fig.~\ref{fig:patio-doors}(b) the percentage deviation of the azimuthally averaged TDS signals from the powder pattern. We see that for the smallest azimuthal averaging range $(10^\circ)$, the averaged TDS differs dramatically (by over 50\%) from a powder signal. As we average over increasingly large regions of reciprocal space, the resulting TDS signal steadily approaches the powder solution; were we able to average over all $\phi$, the TDS pattern would differ from the powder pattern by no more than 5\% over the entire $2\theta$ range measured here. Wide azimuthal detector coverage is therefore integral to the apparent insensitivity of the TDS pattern to crystallographic texture.

The same conclusion can be reached by direct examination of the experimental data. Again, we choose to deliberately restrict the azimuthal averaging region, this time to the domain $\phi\in[0,\Delta\phi]$, where $\Delta\phi=20^\circ,40^\circ$, or $80^\circ$. Since the range of azimuthal angles is already restricted in this case by the finite coverage of the detector, we end up averaging over the blade-shaped regions of the upper detector (Varex 2) shown by Fig.~\ref{fig:expatio-doors}(a). The resulting TDS signals are shown in Fig.~\ref{fig:expatio-doors}(b) along with comparisons with the powder signal and the predictions of our texture-aware TDS model averaged over the same $2\theta$-dependent azimuthal regions. From this figure, two main observations may be made: first, it is clear that the full TDS model provides a stronger match to the data than does the powder model, even for the narrowest azimuthal averaging region where the TDS structure is strongest and the signal noisiest; second, expanding the averaging domain generally pushes the form of the TDS closer to that of an ideal powder.

To summarize, we have shown that the azimuthally averaged thermal diffuse scattering from machine-rolled Cu foils with a moderate texture differs only marginally from that of a perfectly random powder, and indeed is largely insensitive even to the mounting orientation of the foil. Provided the TDS signal can be averaged over x-ray detectors with sufficiently large angular coverage, then, one can reasonably use Warren's one-line analytic expression for (first-order) TDS from a powderlike polycrystal [from Eqs.~(\ref{eq:thermal-textural_decomp}) and (\ref{eq:warren_paskin})] and still expect to obtain an accurate temperature measurement.

\subsection{Influence of grain-sampling statistics\label{sec:grain-statistics}}
\begin{figure}
    \centering
    \includegraphics{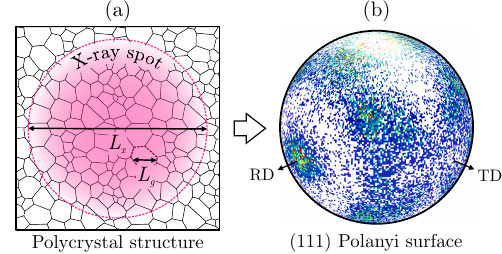}
    \caption{(a) Schematic of an x-ray spot of diameter $L_x$ illuminating a polycrystal with typical grain size $L_g$. (b) Distribution of $\{111\}$ scattering vectors generated from a sample of 5000 grains, showing the rolling and transverse directions (RD and TD respectively).}
    \label{fig:speckly}
\end{figure}

To achieve the intensities needed to launch compression waves of megabar strength ($\sim10^{12}$~Wcm\textsuperscript{$-2$}), long-pulse drive lasers like DiPOLE are typically focused down into a circular spot of 100-500~$\upmu$m diameter. To ensure the high-pressure state being probed is uniform, the x-ray beam is in turn focused into a concentric spot an order of magnitude smaller than the drive spot, i.e., 10-50~$\upmu$m. As shown schematically in Fig.~\ref{fig:speckly}(a), this dimension is not vastly greater than that of the grains being probed. The thickness of the illuminated sample layer, meanwhile, is typically of order 5-25~$\upmu$m; targets much thicker than this cannot be entirely shock-compressed, as the laser pulse supporting the shock typically ends after around 10~ns. Ignoring attenuation effects, the total x-ray--target interaction volume is therefore of order $10^4$~$\upmu$m\textsuperscript{3}. If we make the conservative estimate that the target's grains are at least one micron in size, we should expect the crystallites sampled in a single shot to number in the thousands.

\begin{figure}
    \includegraphics{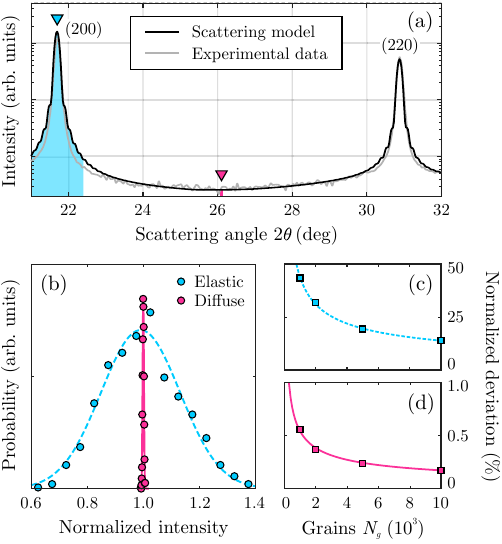}
    \caption{Grain-sampling statistics. (a) Modeled total scattering from a horizontally oriented Cu foil between the (200) and (220) peaks (black) compared with ambient experimental data (gray) on a logarithmic scale. Blue and red triangles at $2\theta$ values of $21.7^\circ$ and $26.2^\circ$ indicate angles at/around which variations in the elastic and diffuse intensities are measured, respectively. (b) Representative intensity distributions obtained from a powder comprising 10\,000 grains, along with fits to normal distributions. Intensities are normalized by the mean of their distribution. (c,d) Scaling of the normalized standard deviation of the intensity distributions with the number of grains $N_g$ for the elastic and diffuse components, respectively. The standard deviations are fitted to an inverse-square-root function.}
    \label{fig:grain-statistics}
\end{figure}

At this kind of sample size, single-shot statistics start to take effect. In reciprocal space, smaller sample sizes manifest as increasingly `spotty' Polanyi surfaces; Fig.~\ref{fig:speckly}(b) shows the predicted structure of $\mathcal{P}_{111}$ for our rolled foils assuming just 5000 grains are sampled. Such spottiness is directly visible in the elastic scattering: the Debye-Scherrer rings in Fig.~\ref{fig:data-summary}(d) -- taken from Cu at 140~GPa probed using a 16-$\upmu$m-wide x-ray spot -- exhibits sharp azimuthal modulations owing to the small number of grains being sampled. Moreover, this azimuthal structure varies visibly from shot-to-shot as the x-ray beam samples different portions of the foil's total texture (the same spot cannot be sampled twice, as dynamic compression obliterates the target). This sampling effect causes the integrated Bragg-peak intensities to change stochastically, rendering measurement of the Debye-Waller factor $M$ via fitting of the elastic scattering alone challenging. Using our model, we can assess whether grain-sampling statistics have a similar effect on the TDS.

To characterize the grain statistics (shown in Fig.~\ref{fig:grain-statistics}), we calculate representative distributions of both the elastic and thermal diffuse scattering for different grain sample sizes. We randomly sample $N_g$ grains from the ODF described in Sec.~\ref{sec:starting-texture} and use our model to evaluate the azimuthally averaged scattering intensity at $26.2^\circ$ [between the (200) and (220) peaks where the TDS dominates], and under the (200) peak centered at $21.7^\circ$ (where the elastic dominates), as indicated in Fig.~\ref{fig:grain-statistics}(a). We do this 1000 times for each sample size $N_g$, and then calculate the normalized distribution of intensities around the two selected scattering angles. We repeat this calculation for sample sizes between 1000 and 10\,000 grains. Azimuthal averages of both the elastic and thermal diffuse scattering are still conducted over the restricted azmiuthal range permitted by the experimental detector configuration described by Fig.~\ref{fig:data-summary}.

To model the elastic scattering, we use the shape function $J$ for spherical coherently diffracting subdomains, whose size is expected to be considerably less than that of the grains themselves \cite{DeAngelis1995,Dubravina2004}. We choose such a domain size as to approximately reproduce linewidth of the experimentally measured Bragg peaks under ambient conditions [as shown by the total scattering prediction in Fig.~\ref{fig:grain-statistics}(a)] and the typical shot-to-shot fluctuations we observe in the ambient Bragg-peak intensities [which, for the (200) peak, are approximately 6\% (25\%) when the targets are probed using a 45~$\upmu$m (16~$\upmu$m) x-ray beam]. 

We predict that for a sample size of $N_g=10\,000$ grains [see Fig.~\ref{fig:grain-statistics}(b)], the intensity of the (200) peak fluctuates by approximately $\pm20\%$, while the TDS between the (200) and (220) peaks deviates by no more than $\pm1\%$ from its average value. As we change the sample size, we find that the standard deviation of the scattering intensity diminishes with $N_g^{-1/2}$ [Fig.~\ref{fig:grain-statistics}(c)]; for grain sample sizes of 10\,000 and above, the statistical error in the TDS is well below the percent level. This figure is considerably smaller than the dominant source of statistical error in the TDS measurement, which -- at time of writing -- comes from normalization of the scattering signal to the incident photon flux, which is measurable with the use of intensity-position monitor (IPM) diodes to within a few percent\cite{Wark2025}. We therefore expect the statistical errors in the TDS resulting from grain-sampling statistics to be essentially negligible, and at least an order of magnitude smaller that that felt by the Bragg peaks.

\subsection{Influence of plasticity-induced texture evolution\label{sec:plasticity}}
To understand the applicability of TDS as a temperature diagnostic to dynamically compressed solids, we must consider not only the statistical variations in (sampled) texture caused by the finite x-ray spot size, but systematic variations in texture caused by plasticity.

\begin{figure}
    \centering
    \includegraphics{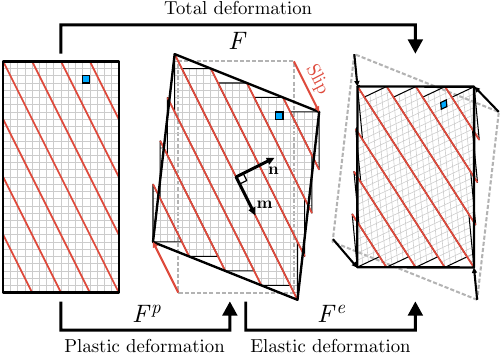}
    \caption{Elastoplastic deformation of a uniaxially compressed single crystal. The total deformation gradient $F = F^eF^p$ comprises a plastic component $F^p$ and an elastic component $F^e$. Plastic deformation is realized via concerted shear motion on a set of slip planes (red), which decreases the crystal's extent along the compression direction but leaves the underlying crystal structure (gray) unchanged. Elastic deformation expresses changes to the crystal structure, including the decrease in volume and the local rotation of the crystal. The shape of a single unit cell (blue) is highlighted for clarity.}
    \label{fig:plasticity}
\end{figure}

When a crystalline sample is uniaxially compressed to stresses beyond its elastic limit, it will deform plastically to relieve the shear stresses accumulated on planes oblique to the loading axis. As indicated in Fig.~\ref{fig:plasticity}, plastic deformation is generally mediated by crystallographic slip, whereby contiguous blocks of the crystal structure slide past one another, in this case allowing the sample to decrease its extent along the compression direction. However, the uniaxial strain imposed upon the sample cannot be realized by plastic deformation alone; the remaining strain is necessarily accommodated via elastic deformation, which generally includes some degree of rotation of the sample's underlying crystal structure. This compression-induced reorientation is grain-specific: a grain's orientation with respect to the loading axis determines the shear stresses felt by its slip systems and thus their activities, meaning each grain experiences a different plastic strain and therefore a different amount (and direction) of rotation. In reciprocal space, plasticity thus manifests as a systematic, conservative redistribution of scattering vectors around the Polanyi surfaces. Plasticity-induced texture changes can be observed directly in the elastic scattering and exploited to obtain information about the plastic deformation mechanisms active under extreme compression conditions\cite{Suggit2012,Wehrenberg2017,Avraam2023}. Does texture evolution imprint similarly on the TDS?

\begin{figure*}
    \centering
    \includegraphics{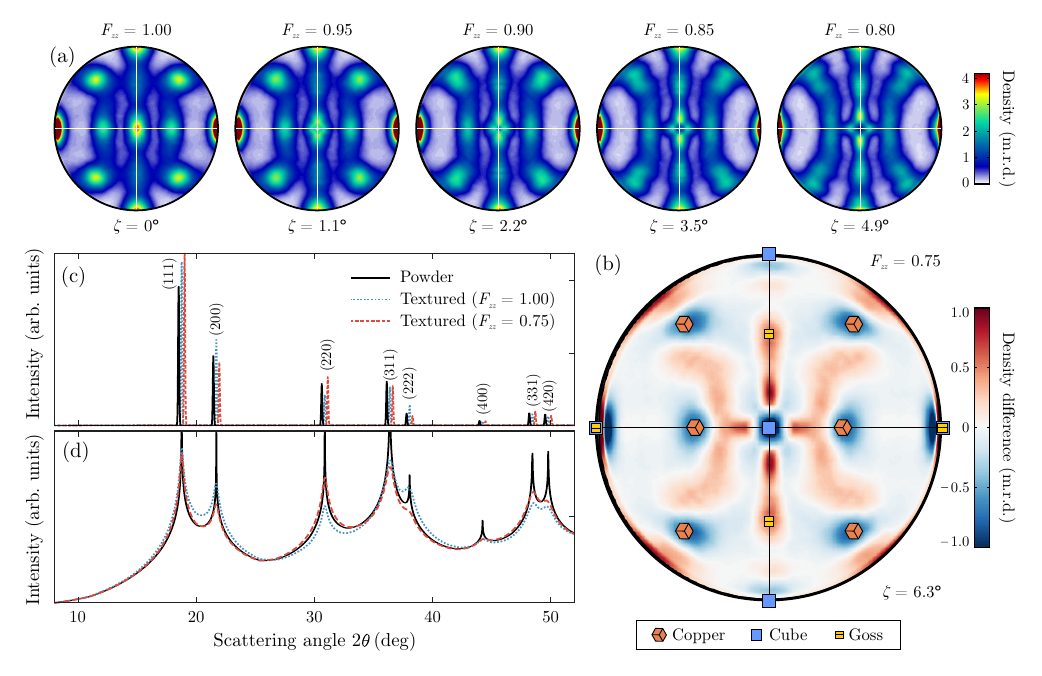}
    \caption{Modeled influence of plasticity-induced texture evolution on elastic and thermal diffuse scattering. (a) Sequence of (200) pole figures for textured Cu polycrystals suffering increasing strain along the compression axis $z$, labeled by the total uniaxial strain $F_{zz}$ and the average grain rotation $\zeta$. (b) Change in the (200) pole figure for a polycrystal compressed to $F_{zz}=0.75$. Overlaid are the high-symmetry copper, cube, and Goss orientations around which the initial orientation distribution function is concentrated. (c,d) Azimuthally averaged elastic and thermal diffuse scattering for both ambient $(F_{zz}=1.00)$ and compressed $(F_{zz}=0.75)$ textured polycrystals, along with the powder signal for comparison. The density change and deviatoric elastic strains accompanying the plastic deformation have been factored out of the calculation. Elastic scattering predictions have been separated by $0.25^\circ$ in $2\theta$ for clarity.}
    \label{fig:plasticity-evolution}
\end{figure*}

We will demonstrate here that, for our moderately textured cubic polycrystals at least, the TDS is only weakly sensitive to texture changes caused by plastic deformation under dynamic compression. To do so, we use a rudimentary model of crystal plasticity for fcc Cu based on the framework described in Refs.~\onlinecite{Heighway2021,Heighway2022}. We will reserve the complete mathematical details of the model for the Supplementary Material, and outline only its key components here.

The central equation connecting plasticity and texture evolution is the multiplicative decomposition of the total deformation gradient $F$ into its elastic and plastic components, $F^e$ and $F^p$, respectively:
\begin{equation}
    F = F^e F^p\ .
\end{equation}
We force the total deformation gradient $F$ to take the form
\begin{equation}
    F = \text{diag}(1,1,F_{zz})\ ,
\end{equation}
reflecting the assumption that every grain in the polycrystal experiences the same uniaxial compression along $z$ (the \textit{Taylor constraint}) by a global factor $F_{zz}=V/V_0$, the total volumetric strain. The plastic deformation $F^p$ is parametrized by the activities on the slip systems, $\{\gamma_\alpha\}$, via the leading-order expression
\begin{equation}
    F^p = I+\sum_\alpha \gamma_\alpha(\mathbf{m}_\alpha\otimes\mathbf{n}_\alpha)\ ,
\end{equation}
where $\mathbf{m}_\alpha$ and $\mathbf{n}_\alpha$ are the direction and plane, respectively, of slip system $\alpha$. The core of our model is estimating the slip-system activities for each grain, allowing us to calculate its elastic deformation gradient $F^e=FF^{p-1}$ and thus the change in size, shape, and orientation of its crystal structure. We do this by first estimating the \emph{relative} activity of each grain's slip systems via the resolved shear stresses they experience (given the grain's orientation, its elastic moduli, and the total volumetric strain) before computing the \emph{absolute} activity on each system by finding the accessible strain state with the least deviatoric elastic strain.

In general, the outcome of this calculation is an elastic strain state including both a density increase (as enforced by the compression factor $F_{zz}$) and elastic shear strains that warp the unit cell. To simplify the analysis, we will dispense with both of these factors and extract only the purely rotational part of the elastic deformation, $R$, via the polar decomposition:
\begin{equation}
    F^e = RU^e\ .
\end{equation}
In so doing, we essentially allow each grain's scattering vectors only to move around their original spherical Polanyi surfaces, without those surfaces changing shape or size. We do this to maintain compatibility with the overall structure of our TDS model and to make comparison between signals from polycrystals with different densities easier; here, we are interested not in the fact that the scattering signal is everywhere pushed out to higher $q$, but in the changes in crystallographic texture accompanying the compression. We further ignore the possibility of deformation twinning, which would bring about nanograins with new orientations. We should stress that this simplistic model is not intended to yield a quantiatively accurate prediction of the grain-dependent rotation -- something which would be better obtained with a spatiotemporally resolved, dynamic crystal plasticity model \cite{Avraam2023} -- but merely to actively rotate each grain about a unique axis by an angle broadly consistent with that expected under shock (no more than $10^\circ$ at 100~GPa).

To ensure that the systematic plasticity-induced texture evolution dominates over statistical variations caused by finite grain sampling, we simulate polycrystals comprising $10^6$ grains; according to the sampling statistics shown in Fig.~\ref{fig:grain-statistics}, such a grain population reduces random fluctuation in Bragg-peak intensities around $1\%$, while that of the TDS becomes entirely negligible.

Fig.~\ref{fig:plasticity-evolution}(a) shows the systematic change in the distribution of the $\{200\}$ scattering planes for uniaxial strains between 0 and 0.20. The change in Polanyi density is made more apparent by Fig.~\ref{fig:plasticity-evolution}(b), which shows the difference between the angular density of $\{200\}$ vectors for an ambient polycrystal and one compressed to $F_{zz}=0.75$ (effecting the same amount of plastic deformation as shock-compression to around 90~GPa). Our simple model predicts that both the copper and cube orientations are unstable: grains which sit close to these points in orientation space rotate away from these ideal orientations, leading to the formation of `limbs' spanning the previously weakly occupied orientation states between them. In particular, grains with the cube orientation undergo \textit{conjugate slip}, rotating about one of the $\langle100\rangle$ directions orthogonal to the compression direction such that one of the $\langle110\rangle$ directions rotates towards $z$, in agreement with both dynamic-compression experiments on \cite{Suggit2012} and molecular dynamics simulations of \cite{Heighway2022} [001]-oriented Cu. By contrast, the Goss orientation becomes more prevalent, consistent with the expected stability of the fcc $[110]$ direction under compression \cite{Khan1995}. For a compression of $F_{zz}=0.80$, the average grain rotation is just under $5^\circ$, in agreement with measurements from dynamic-compression experiments of [001]-oriented Cu \cite{Suggit2012}.

When we examine the plasticity-induced changes to the azimuthally averaged elastic and TDS patterns [Fig.~\ref{fig:plasticity-evolution}(c,d)], we observe strikingly different behavior. The changes in the Bragg-peak intensities are relatively dramatic, approaching 50\% for the (220) and (222) reflections. By contrast, the TDS in the regions between widely separated Bragg peaks (i.e., those TDS-dominated regions whence we would derive a temperature in experiment) changes by no more than 10\% in any interval; again, we observe the characteristically delocalized nature of the TDS `softening' its sensitivity to texture. This physics is illustrated most vividly by the practically unchanged TDS signal between the (200) and (220) Bragg peaks. Here, the dominant contributions to the TDS are from scattering vectors on the two nearest Polanyi surfaces, $\mathcal{P}_{200}$ and $\mathcal{P}_{220}$. The number of scattering vectors suitably situated to contribute to the TDS can be estimated by proxy from the intensity TDS directly beneath the Bragg peaks, which we see diminishes for the (200) and grows for the (220). It transpires that increasing TDS from the (220) and decreasing TDS from the (200) vectors compensate for one another almost exactly, accounting for the observation that the TDS is essentially stationary with respect to texture evolution midway between the peaks.

It is also striking that, upon compression, the structure of the TDS from the polycrystal converges to that of a powder. This can be understood with reference to the pole plots in Fig.~\ref{fig:plasticity-evolution}(a), which demonstrate that the overall effect of plastic deformation on these moderately textured polycrystals is to diversify the range of grain orientations present, and thus reduce the additional structure in the TDS caused by the anisotropy of the Polanyi densities. We should stress, however, that this is \emph{not} the behavior we observe in experiment: in Fig.~\ref{fig:data-summary}(d), we see that the quality of Warren's powder fit to the data is actually poorer for shock-compressed targets. We speculate that this additional structure in the TDS is due not to texture changes but to growing anharmonicity along the Hugoniot, which is unaccounted for in the current form of our model. Nevertheless, we conclude that slip-mediated plastic deformation experienced by a dynamically compressed rolled foil is unlikely to significantly change the form of its TDS pattern, and may in fact drive that pattern closer yet towards that of an ideal powder if the polycrystal's starting texture is sufficiently strong.

\section{Discussion\label{sec:discussion}}

Dynamic-compression experiments are soon to enter an era of unprecedented repetition rates. The advent of diode-pumped, cryogenically cooled, long-pulse drive lasers -- exemplified by the DiPOLE 100-X laser system recently commissioned at the European XFEL \cite{Gorman2024}, which is capable of firing at 10~Hz -- means the number of data points accrued in a single experiment will soon be measured not in hundreds but thousands.

Such high-throughput experiments demand a huge number of targets. To meet these demands, it is often most practicable to source multiple batches of samples: some may be bought `off-the-shelf' from commercial vendors that produce thin foils \textit{en masse} via machine rolling, imparting a characteristic texture over which the user has little control; others may be procured via physical vapor deposition (PVD), the outcome of which can vary considerably unless the preparation `recipe' is specified very precisely. Consequently, the target make-up of a multi-day dynamic-compression experiment conducted at a rep-rate facility often comprises a highly heterogeneous mixture of multiple batches of sample with subtly (or perhaps dramatically) different crystallographic textures that are difficult to track and sometimes impossible to reproduce in subsequent campaigns.

The overall picture of thermal diffuse scattering that emerges from the foregoing modeling is one of remarkable insensitivity to such textural variations. We have shown that, for these moderately textured Cu polycrystals, the azimuthally averaged TDS (1) is largely insensitive to the sample mounting orientation, (2) fluctuates shot-to-shot at the sub-percent level even when only a few thousand grains are sampled, (3) changes by no more than 10\% due to systematic, compression-induced texture changes, and (4) in all cases strongly resembles the TDS from an ideal random powder. The last of these observations is particularly powerful, as it means one can justifiably use Warren's one-line analytic expression for the (first-order) TDS to treat the experimental data and still expect to obtain a highly accurate measurement of the Debye-Waller factor (and hence the temperature), without having to appeal to texture-characterization techniques and the relatively extensive model described in Sec.~\ref{sec:maths}.

This behavior contrasts starkly with that of the elastic scattering: individual Bragg-peak intensities may vary by around 20\% due to sampling statistics and by up to 50\% due to plasticity. In other words, the intensity of any given Bragg peak is modulated not only by the Debye-Waller factor, but by a peak-specific, texture-dependent correction [encoded by the $\sigma_{hkl}$ term in Eq.~(\ref{eq:diffraction-to-texture})]. Failing to account for this factor and naively calculating $2M$ via the logarithm of the Bragg-peak intensity could result in inferred temperatures that are inaccurate by several tens of percent. Thermal diffuse scattering is therefore preferable as a temperature diagnostic if one cannot practicably calculate the ODF of the compressed target \textit{in situ} (due to insufficient detector coverage, for example).

We have shown that the insensitivity of TDS to texture is contingent on its being azimuthally averaged over a sufficiently wide range of azimuthal angles. Wide coverage of $\mathbf{q}$-space is readily achievable in a standard diffraction experiment measuring TDS, but the same is generally untrue of IXS-based experiments in which that same TDS is spectrally resolved; the energy-dispersive elements employed typically sample only a small area of the Ewald sphere (and, by extension, the Polanyi surfaces). In forward scattering (low $M$), the limited textural coverage will impart a nontrivial structure to the Stokes and anti-Stokes peaks that is sensitive to the sample's orientation. In principle, the ratio of the \emph{total} areas under these peaks still depends only on the system's temperature, but interpreting the inelastic peaks may still prove difficult in practice if they partially overlap a quasielastic peak at zero energy transfer (due to crystal defects, for example \cite{Karnbach2021}) or if the photostatistics are poor. The texture sensitivity may be overcome by moving to a backscattering (high $M$) geometry, in which limit the meV-IXS spectrum collapses to a Gaussian distribution whose height depends on texture, but whose width varies with temperature only. In this sense, IXS spectra collected in the single-particle regime may also be considered partially `texture-robust'. In principle, it would be simple to augment the mathematical framework described here and calculate \emph{spectrally resolved} scattering patterns [i.e., $S(\mathbf{q},\omega)$] in the single-phonon scattering regime at least, allowing us to model in detail the influence of crystallographic texture on meV-IXS spectra.

We have further demonstrated that for moderately textured polycrystals whose TDS pattern differs marginally from that of a powder, the small `shortfall' can be successfully accounted by for our model's inclusion of the ODF [Fig.~\ref{fig:master}(d,h)], allowing us to extend the domain of applicability of TDS-based thermometry. However, there must exist a level of texture beyond which our model will fail. Among its simplifying assumptions are (1) that the Brillouin zone is spherical and populated by phonons with a linear, isotropic dispersion relation and (2) that higher-order TDS arising from multi-phonon scattering events is largely structureless [Eq.~(\ref{eq:borie})]. These approximations will surely break down in the limit of the most strongly textured specimen possible, i.e., a single crystal, for which $S_{\rm{TD}}(\mathbf{q})$ is known to exhibit the same strong anisotropy as the crystal structure itself. The structure of the single-crystal TDS is `washed out' if the polycrystal's distribution of grain orientations is sufficiently diverse -- as it is for our commercial foils -- but there will exist a degree of texture beyond which such structure can no longer be ignored, and a model of single-crystal-level TDS more elaborate than that provided by the Debye picture would be required.

Indeed, there is a general need to explore and push the limits of applicability of TDS modeling. The principle of extracting temperatures from dynamically compressed matter via femtosecond thermal diffuse scattering measurements has thus far been realized in only one scenario: shock-compressed, polycrystalline, fcc copper. If TDS as a temperature diagnostic is to be widely adopted, we must be able to model the diffuse scattering from samples with a diversity of crystal structures compressed via a variety of loading paths, be they mono- or polycrystalline. Intense investigation into thermal-diffuse-scattering theory and its application to femtosecond scattering from dynamically compressed matter thus represents an immediate and important research opportunity.

\section{Conclusion\label{sec:conclusion}}

We have developed a texture-aware model of x-ray thermal diffuse scattering (TDS) from cubic polycrystals. We have shown that our model successfully reproduces the TDS patterns measured from moderately textured Cu rolled foils, and with greater accuracy than does Warren's classic analytic model of TDS from a perfectly random powder. Our model further predicts that, with sufficient detector coverage, the azimuthally averaged TDS pattern is remarkably insensitive to the details of the sample's underlying crystallographic texture, varying statistically at the percent level due to the sampling of finitely many grains and systematically by no more than 10\% due to compression-induced crystal plasticity. These results endorse TDS as a robust temperature diagnostic for dynamically compressed solids whose potential we have only begun to explore.

\section*{Supplementary Material}
See the Supplementary Material for details of how we validated the texture-integration scheme described in Sec.~\ref{sec:polycrystal-scattering} and the structure of the plasticity model outlined in Sec.~\ref{sec:plasticity}.

\begin{acknowledgments}

PGH and JSW gratefully acknowledge support from EPSRC under research grant 	EP/X031624/1. DJP and TS appreciate support from AWE via the Oxford Centre for High Energy Density Science (OxCHEDS).

We acknowledge the European XFEL in Schenefeld, Germany, for provision of X-ray free electron laser beam time at the Scientific Instrument HED (High Energy Density Science) and would like to thank the staff for their assistance. The authors are indebted to the Helmholtz International Beamline for Extreme Fields (HIBEF) user consortium for the provision of instrumentation and staff that enabled this experiment. The data is available upon reasonable request [doi: 10.22003/XFEL.EU-DATA-002740-00].

We acknowledge support for the provision of the DiPOLE laser from the UK STFC and EPSRC under grant numbers EP/M000508/1 and EP/L022591/1.

We acknowledge DESY (Hamburg, Germany), a member of the Helmholtz Association HGF, for the provision of experimental facilities. Parts of this research were carried out at PETRA III (beamline P02.2).

Part of this work was performed under the auspices of the U.S.\ Department of Energy by Lawrence Livermore National Laboratory under Contract No.\ DE-AC52-07NA27344 and was  supported by the Laboratory Directed Research and Development Program at LLNL (Project No.\ 21-ERD-032).

Part of this work was performed under the auspices of the U.S.\ Department of Energy through the Los Alamos National Laboratory, operated by Triad National Security, LLC, for the National Nuclear Security Administration (Contract No.\ 89233218CNA000001).

Research presented in this article was supported by the Department of Energy, Laboratory Directed Research and Development program at Los Alamos National Laboratory under Project No.\ 20190643DR and at SLAC National Accelerator Laboratory, under Contract No.\ DE-AC02-76SF00515.

This work was supported by Grants No.\ EP/S022155/1 (MIM, MJD) EP/S023585/1 (AH, LA) and EP/S025065/1 (JSW) from the UK Engineering and Physical Sciences Research Council. JDM is grateful to AWE for the award of CASE Studentship P030463429.

EEM and AD were supported by the UK Research and Innovation Future Leaders Fellowship (MR/W008211/1) awarded to EEM. 

DE and DS from Univ.\ de Valencia thank the financial support by the Spanish Ministerio de Ciencia e Innovación (MICINN) and the Agencia Estatal de Investigación (MCIN/AEI/10.13039/501100011033) under grants 
PID2021-125518NB-I00
and PID2022-138076NB-C41 (cofinanced by EU FEDER funds), and by the Generalitat Valenciana under grants CIPROM/2021/075, CIAICO/2021/241 and MFA/2022/007 (funded by Next Generation EU PRTR-C17.I1).

The work at USF has been supported by Lawrence Livermore National Laboratory's Academic Collaborative Team award, DOE/NNSA (Award Nos.\ DE-NA-0004089 and DE-NA0004234) and DOE/FES (Award No.\ DE-SC0024640), the National Science Foundation (Award No.\ 2421937), National Aeronautics and Space Administration (24-XRP24\_2-0176).

NJH and AG were supported by the DOE Office of Science, Fusion Energy Science under FWP 100182. This material is based upon work supported by the Department of Energy National Nuclear Security Administration under Award Number DE-NA0003856.

YL is grateful for the support from the Leader Researcher program (NRF-2018R1A3B1052042) of the Korean Ministry of Science and ICT (MSIT).

KA, KB, ZK, HPL, RR and TT thank the DFG for support within the Research Unit FOR 2440. KB acknowledges financial support via project AP262/2-2 (Project No.\ 280637173).

BM and RSM acknowledge funding from the European Research Council (ERC) under the European Union’s Horizon 2020 research and innovation programme (Grant agreement No.\ 101002868).

GWC and T-AS recognize support from NSF Physics Frontier Center Award No.\ PHY-2020249 and support by the U.S.\ Department of Energy National Nuclear Security Administration under Award No.\ DE-NA0003856, the University of Rochester, and the New York State
Energy Research and Development Authority.

SM, HG, and JC are funded by the European Union (ERC, HotCores, Grant No.\ 101054994). Views and opinions expressed are however those of the author(s) only and do not necessarily reflect those of the European Union or
the European Research Council. Neither the European Union nor the granting authority can be held responsible for them. 

The work of DK, DR, JR, and MS was supported by Deutsche Forschungsgemeinschaft (DFG—German Research Foundation) Project No.\ 505630685.

SP acknowledges support from the GotoXFEL 2023 AAP from CNRS.

\end{acknowledgments}

\section*{Author declarations}
The authors have no conflicts or competing interests to disclose.

\section*{Data availability}
The x-ray diffraction data obtained from the experiment at EuXFEL described in this article are available at \href{https://doi.org/10.22003/XFEL.EU-DATA-002740-00}{doi:10.22003/XFEL.EU-DATA-002740-00}.

\section*{References}
%

\end{document}


\title{Supplementary Material: X-ray thermal diffuse scattering as a texture-robust temperature diagnostic for dynamically compressed solids}

\author{P.G. Heighway~\orcidlink{0000-0001-6221-0650}}\email{patrick.heighway@physics.ox.ac.uk}
\affiliation{Department of Physics, Clarendon Laboratory, University of Oxford, Parks Road, Oxford OX1 3PU, UK\looseness=-1}

\author{D.J. Peake\orcidlink{0000-0002-5992-6954}}
\affiliation{Department of Physics, Clarendon Laboratory, University of Oxford, Parks Road, Oxford OX1 3PU, UK\looseness=-1}

\author{T. Stevens~\orcidlink{0009-0006-8355-3509}}
\affiliation{Department of Physics, Clarendon Laboratory, University of Oxford, Parks Road, Oxford OX1 3PU, UK\looseness=-1}

\author{J.S. Wark~\orcidlink{0000-0003-3055-3223}}
\affiliation{Department of Physics, Clarendon Laboratory, University of Oxford, Parks Road, Oxford OX1 3PU, UK\looseness=-1}


\author{B. Albertazzi}\affiliation{Ecole Polytechnique, Palaiseau, Laboratoire pour l'utilisation des lasers intenses (LULI), CNRS UMR 7605 Route de Saclay 91128 PALAISEAU Cedex, France\looseness=-1}

\author{S.J. Ali~\orcidlink{0000-0003-1823-3788}}\affiliation{Lawrence Livermore National Laboratory, Livermore, CA 94550, USA\looseness=-1}

\author{L. Antonelli~\orcidlink{0000-0003-0694-948X}}\affiliation{University of York, School of Physics, Engineering and Technology, Heslington York YO10 5DD, UK\looseness=-1}

\author{M.R. Armstrong~\orcidlink{0000-0003-2375-1491}}\affiliation{Lawrence Livermore National Laboratory, Livermore, CA 94550, USA\looseness=-1}

\author{C. Baehtz~\orcidlink{0000-0003-1480-511X}}\affiliation{Helmholtz-Zentrum Dresden-Rossendorf (HZDR), Bautzner Landstra{\ss}e 400, 01328 Dresden, Germany\looseness=-1}

\author{O.B. Ball~\orcidlink{0000-0002-5215-0153}}\affiliation{SUPA, School of Physics and Astronomy, and Centre for Science at Extreme Conditions, The University of Edinburgh, Edinburgh EH9 3FD, UK\looseness=-1}

% ...

\date{\today}

\begin{abstract}
This Supplementary Material details how we verified the accuracy of the texture integral required to calculate thermal diffuse scattering from an arbitrarily textured polycrystal, and outlines the structure of our simple crystal plasticity model.
\end{abstract}

\maketitle

\section{Accuracy of the texture integral\label{sec:texture-integral}}
In our x-ray scattering model, the first-order thermal diffuse scattering (TDS) from a polycrystal comprising $N_g$ grains, each containing $N_a$ atoms, is expressed as
\begin{equation}
    S_1(\mathbf{q}) = N_a N_g f^2 2Me^{-2M} \underbrace{\sum_{hkl}\iint\limits_{\mathcal{P}_{hkl}} d\Omega\,\hat{\sigma}_{hkl}(\mathbf{P})W(\mathbf{q}-\mathbf{P})}_{C_1(\mathbf{q})}\ ,
\end{equation}
where $f(\mathbf{q})$ is the atomic form factor, $M(\mathbf{q})$ the Debye-Waller factor, $\hat{\sigma}_{hkl}(\mathbf{P})$ the number of scattering vectors per solid angle per grain at point $\mathbf{P}$ on Polanyi surface $\mathcal{P}_{hkl}$, and where $W$ (the Warren kernel) takes the form
\begin{equation}
    W(\mathbf{k}) = \begin{cases}
    \frac{1}{3}\frac{q_B^2}{k^2} & \text{for}\quad 0<k\le q_B\ , \\
    0 & \text{for}\quad k>q_B\ ,
    \end{cases}
\end{equation}
with $q_B$ being the radius of the (spherical) Brillouin zone. The simple form of the Warren kernel $W$ makes it possible to evaluate the geometric factor $C_1$ analytically in certain high-symmetry (and experimentally relevant) scenarios. Direct comparison with such analytic solutions allows us to verify the accuracy of the texture-dependent component of our TDS model.

The case originally treated by Warren \cite{Warren1953} was that of a perfectly isotropic powder. In this scenario, every Polanyi surface is completely and uniformly occupied by reciprocal lattice vectors of constant angular density
\begin{equation}
    \hat{\sigma}_{hkl} = \frac{j_{hkl}}{4\pi}\ .
\end{equation}
The contribution of a single Polanyi surface $\mathcal{P}_{hkl}$ to the texture integral $C_1$ reduces to the simple form first derived by Warren:
\begin{equation}
    C_1^{(hkl)}(\mathbf{q}) = \frac{1}{6}j_{hkl}\frac{q_B^2}{qG_{hkl}}\ln\left(\frac{q_B}{|q-G_{hkl}|}\right)\ .
\end{equation}
The expression above holds only for $|q-G_{hkl}| < q_B$, which is to say, for $q$-points within distance $q_B$ of $\mathcal{P}_{hkl}$; otherwise, the integral vanishes.

\begin{figure}
    \includegraphics{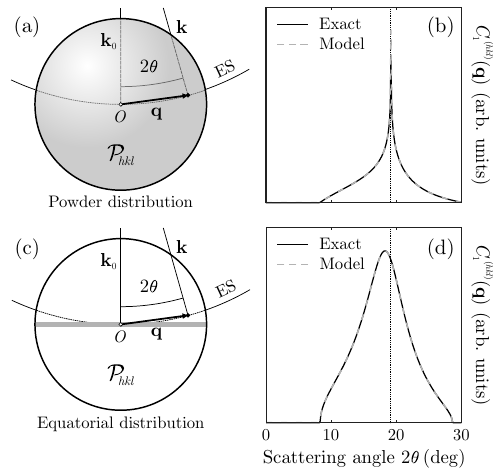}
    \caption{Comparison between exact (analytic) and modeled (numerical) texture integral $C_1^{(hkl)}$ for select high-symmetry polycrystals. (a)~Reciprocal-space schematic showing the intersection between a uniformly occupied Polanyi surface $\mathcal{P}_{hkl}$ and the Ewald sphere (ES) for an ideal powder sample. (b)~Exact and modeled forms of $C_1^{(111)}$ for an 18~keV x-ray source and a cubic lattice constant $a$ of 3.62~\AA\ ($2\theta_{111} = 19^\circ$). (c,d)~Equivalent figures for a Polanyi surface occupied by scattering vectors restricted to a one-dimensional equatorial locus.}
    \label{fig:integration}
\end{figure}

We show in Fig.~\ref{fig:integration}(b) the form of this integral calculated both analytically and by numerical integration over a uniform Polanyi surface. For illustration, we choose the $\{111\}$ scattering vectors from an fcc Cu polycrystal ($a$ = 3.62~\AA) illuminated by an 18~keV x-ray source. The model reproduces the classic broad peak predicted by the Warren model, whose maximum is a signature logarithmic divergence located where the Polanyi surface and Ewald sphere meet, i.e., at exactly the same scattering angle $2\theta_{hkl}$ that the Bragg (elastic) peaks from that same Polanyi surface appear.

The second high-symmetry case we consider is one in which the Polanyi density is concentrated into a single circular locus embedded in the Polanyi surface. Reciprocal-space `orbits' of this kind are relevant for fiber-textured samples whose macroscopic texture is invariant when rotated around a global axis. Such samples are not merely of academic interest: they can typically be procured via physical vapor deposition, and their sparse reciprocal-space structure allows them to offer unique insights into microstructure dynamics under dynamic compression \cite{Wehrenberg2017,Sliwa2018,Heighway2019,Avraam2023}.

For convenience, we consider a single uniform locus of scattering vectors that occupy the `equator' of the Polanyi surface, as pictured in Fig.~\ref{fig:integration}(c). An equatorial locus of $\{111\}$ scattering vectors could be encountered, for example, in a polycrystalline target with a $[110]$ fiber axis whose direction coincided with that of the incident x-ray beam. The cylindrical symmetry of the equatorial distribution means that, like the powder distribution, it yields a scattering signal that is insensitive to the azimuthal angle $\phi$ measured around the beam direction.

To perform the textural integral, we convert the two-dimensional integral over solid angle to a one-dimensional integral over angle $(\int d\Omega \to \int d\varphi)$ and write the Polanyi density (now expressed per unit angle) as
\begin{equation}
    \hat{\sigma}_{hkl} = \frac{j_{hkl}'}{2\pi}\ .
\end{equation}
Here, $j_{hkl}'$ is the multiplicity of the reciprocal lattice vectors within the locus, which is less than the total multiplicity $j_{hkl}$; for the $[110]$ fiber axis, for instance, $j_{111}' = 4$, while $j_{111}=8$. The textural integral can still be evaluated analytically, but it takes a relatively complex form:
\begin{equation}
\begin{split}
    C_1^{(hkl)}(\mathbf{q}) &= \frac{1}{3\pi}\frac{q_B^2}{q_T G_{hkl}}j_{hkl}' \\
    &\times \frac{1}{\sqrt{\alpha^2 - 1}}\arctan\left[\sqrt{\frac{(\alpha+1)(1-\alpha+\beta)}{(\alpha-1)(1+\alpha-\beta)}}\right]\ ,
\end{split}
\end{equation}
where the dimensionless quantities $\alpha$ and $\beta$ are given by
\begin{align}
    \alpha(\mathbf{q}) &= \frac{q^2+G_{hkl}^2}{2q_T G_{hkl}}\ , \\
    \beta(\mathbf{q}) &= \frac{q_B^2}{2q_T G_{hkl}}\ ,
\end{align}
with $q_T$ being the projection of $\mathbf{q}$ onto the equatorial plane. The expression above holds in the region of reciprocal space for which $\alpha-\beta<1$; otherwise, the integral vanishes.

We show the analytic and modeled integrals for the equatorial distribution in Fig.~\ref{fig:integration}(d). In contrast to the powder distribution, the locus of intersection between the Ewald and Polanyi spheres is devoid of scattering vectors, meaning we do not see a logarithmic divergence, but a flatter, finite maximum at a marginally lower scattering angle of $18.3^\circ$. This corresponds approximately to the points on the Ewald sphere closest to the ring of reciprocal-space intensity, which are those satisfying $2\theta = \arctan(G_{hkl}/k)=18.5^\circ$.

In both cases, the known analytic solution is reproduced correctly by the numerical integral.

\section{Plasticity model\label{sec:plasticity-model}}
To model qualitatively the texture evolution caused by crystal plasticity under uniaxial compression conditions, we use a rudimentary model based largely on the framework described in Refs.~\onlinecite{Heighway2021,Heighway2022}. The essence of the model is to estimate the proportion in which each grain's slip systems become active, and then to identify the kinematically accessible state nearest the hydrostat.

We start from the elastoplastic decomposition of the total gradient $F$ into its elastic component $F^e$ and its plastic component $F^p$:
\begin{equation}
    F = F^e F^p\ .
\end{equation}
On a global level, the total deformation gradient imposed upon a uniaxially, dynamically compressed target is
\begin{equation}
    F' = \text{diag}(1,1,F_{zz}')\ ,\label{eq:lateral-confinement}
\end{equation}
where the prime signifies we are working with target or laboratory coordinates. The equation above expresses the fact that the target contracts by factor $F_{zz}'$ along the compression axis, $z$, but suffers no change to its transverse dimensions. This follows from the extremely rapid nature of dynamic compression experiments, over whose duration ($\sim10$~ns) the central driven part of the target cannot expand laterally via the Poisson effect due to its being confined by similarly pressurized surrounding material. In our model, we further assume that the conditions of lateral confinement also apply \emph{locally}, by which we mean that $F'$ takes the diagonal form given by Eq.~(\ref{eq:lateral-confinement}) for every grain in the polycrystal. Enforcing homogeneity of $F'$ means ignoring the forces exerted on each grain by its neighbors that would in reality allow them to expand or contract cooperatively in the transverse directions \cite{Heighway2019b}. The faithful treatment of these effects would call for a model far more complex than what we require for the purposes of the present study, and so we choose for convenience to assume $F'$ is diagonal throughout the polycrystal.

To estimate the plastic deformation experienced by a particular grain, we first calculate the proportion in which its slip systems become active based on the resolved shear stresses they experience. We imagine that a grain immediately behind the shock front that has yet to undergo plastic deformation initially finds itself in a state of purely elastic deformation identical to the globally (and locally) imposed total deformation:
\begin{equation}
    F^{e\prime}_0 = \text{diag}(1,1,F_{zz}')\ .
\end{equation}
We calculate this initial elastic deformation in the frame of the grain of interest via the usual transformation
\begin{equation}
    F^e_0 = RF_0^{e\prime}R^T\ ,
\end{equation}
where the matrix converting from the target frame to the grain frame, $R$, is derived from its Euler angles. We then calculate the initial stress state in the grain frame, $\sigma$, using standard elasticity theory. Given the Green strain
\begin{equation}
    e = \frac{1}{2}\left[(F_0^e)^TF_0^e-I\right]\ ,
\end{equation}
we use Hooke's law,
\begin{equation}
    \sigma = ce\ ,
\end{equation}
to compute the attendant stresses using high-pressure elastic constants calculated with density functional theory (DFT) by Bai and Liu \cite{BaiLiGang2020}; the appropriate pressure $p$ at which to evaluate $c$ is estimated by finding the pressure along the 300~K isotherm for volume $V=F_{zz}V_0$. The fourth-rank elasticity tensor $c$ is calculated using the eigentensor representation by Mehrabadi and Cowin \cite{MehrabadiCowin1990}.

With the grain's known stress state $\sigma$, we evaluate the resolved shear stresses $\{\tau_\alpha\}$ acting on its slip systems, which we enumerate with the index $\alpha$. Given each system's slip direction $\mathbf{m}_\alpha\in\langle110\rangle$ and slip plane normal $\mathbf{n}_\alpha\in\{111\}$, the resolved shear stress it experiences is
\begin{equation}
    \tau_\alpha = \mathbf{n}_\alpha^T\sigma\mathbf{m}_\alpha\ .
\end{equation}
We then use a phenomenological flow rule to estimate the glide $\gamma_\alpha$ undergone on slip system $\alpha$:
\begin{equation}
    \gamma_\alpha = a\,\text{sgn}(\tau_\alpha)|\tau_\alpha|^b\ .
\end{equation}
Here, $a$ controls the \emph{absolute} amount of activity on the slip systems, while the exponent $b$ governs the \emph{relative} activity of the slip systems; the greater $b$, the less equitably is glide distributed.

The plastic deformation experienced by the grain is, to leading order, related to the slip-system activities by
\begin{equation}
    F^p = I+\sum_\alpha \gamma_\alpha(\mathbf{m}_\alpha\otimes\mathbf{n}_\alpha)\ .
\end{equation}
For fixed $\{\tau_\alpha\}$ and $b$, the grain's allowed elastic deformation states $F^e = FF^{p-1}$ are confined to a one-dimensional locus parametrized by the variable $a$. To assign the grain its state, we identify the value of $a$ yielding the state nearest the hydrostat, as measured by the scalar, frame-invariant von-Mises elastic shear strain $\varepsilon^e_{\rm{vms}}\ge0$. Given the elastic deformation gradient $F^e$, we factor out the associated rigid-body rotation $R^e$ via the polar decomposition
\begin{equation}
    F^e = R^eU^e
\end{equation}
before calculating the deviatoric part of the right stretch tensor $U^e$ via
\begin{equation}
    \varepsilon^e = U^e - \left(\frac{1}{3}\text{Tr}\,U^e\right)I
\end{equation}
and then calculating the von-Mises elastic shear strain
\begin{equation}
    \varepsilon_{\rm{vms}}^e = \sqrt{\frac{2}{3}(\varepsilon^e:\varepsilon^e)}\ ,
\end{equation}
where $:$ represents the double dot product. In this context, $\varepsilon^e_{\rm{vms}}$ (and indeed $F^e$ and $F^p$) can be considered one-dimensional functions of the single variable $a$, which expresses the total amount of plastic activity for the grain under consideration. The von-Mises strain $\varepsilon_{\rm{vms}}^e(a)$ is convex: before plastic deformation, $\varepsilon_{\rm{vms}}^e$ takes a finite value associated with the highly deviatoric initial elastic strain state $F^e(a=0)=F^e_0$; letting $a$ increase will at first allow the grain to relax its shear strains towards their minimal accessible values; allowing $a$ to increase too far results in an `overshoot', and the shear strains will start to build once more. The elastic deformation state $F_g^e$ we assign to the grain is that which minimizes $\varepsilon_{\rm{vms}}^e$:
\begin{subequations}
    \begin{align}
        F^e_g &= F^e(a_{\rm{min}})\ , \\
        a_{\rm{min}} &= \argmin_{a\,\in\,\mathbb{R}} \varepsilon_{\rm{vms}}^e(a)\ .
    \end{align}
\end{subequations}

Finally, we re-express the grain's elastic deformation in the target frame using the inverse transformation
\begin{equation}
    F_g^{e\prime} = R^TF_g^eR\ .
\end{equation}
We can hence transform the grain's scattering vectors as required to update the full target's reciprocal-space structure and thus recalculate its scattering pattern. In general, the reciprocal lattice vector $\mathbf{G}$ would be actively transformed according to
\begin{equation}
    \mathbf{G}\to[(F_g^e)^T]^{-1}\mathbf{G}\ ;
\end{equation}
in the Main Article, we factor out both the volume change and elastic shear strains associated with $F_g^e$ by calculating its rotational part using a polar decomposition.

Our plasticity model has only one free parameter: $b$. For the present study, we choose $b=10$, as this predicts an average grain rotation of approximately $5^\circ$ at a compression of $F_{zz}=0.80$, in agreement with \textit{in situ} rotation measurements made by Suggit \textit{et al.}~on shock-compressed, $[001]$-oriented Cu single crystals\cite{Suggit2012}.

\section*{References}
%